\documentclass[useAMS,usenatbib,usegraphicx]{mn2e}

\title[Galaxy Structure in Stellar Mass and Light at $z < 1.0$]{The Structures of Distant Galaxies V: The Evolution of Galaxy Structure in Stellar Mass at $z < 1$} 
\author[M. M. Lanyon-Foster, C. J. Conselice, M. R. Merrifield]{M. M. Lanyon-Foster$^{1}$, C. J. Conselice$^{1}$\thanks{E-mail:
conselice@nottingham.ac.uk}, M. R. Merrifield$^{1}$ \\
$^{1}$University of Nottingham, School of Physics \& Astronomy, Nottingham, NG7 2RD UK}
\def\deg{$^{\circ}\,$}

\def\etal {{et~al.$\,$}}
\def\deg{$^{\circ}\,$}

\def\casgm20{CAS-G-M$_{20}\,$}
\def\m20{M$_{20}\,$}

\def\zband{$z_{850}\,$}

\begin{document}

\date{Accepted ; Received ; in original form}

\pagerange{\pageref{firstpage}--\pageref{lastpage}} \pubyear{2009}

\maketitle

\label{firstpage}
 
\begin{abstract}

Galaxy structure and morphology is nearly always studied using the 
light originating 
from stars, however ideally one is interested in measuring structure 
using the stellar mass distribution. Not only does stellar mass trace out 
the underlying distribution of matter it also minimises the effects of 
star formation and dust on the appearance and structure of a galaxy.    
We present in this paper a study of the stellar mass distributions and 
structures of galaxies at $z < 1$ as found within the GOODS fields.
We use pixel by pixel K-corrections to construct stellar
mass and mass-to-light ratio maps of 560 galaxies of known morphology at
magnitudes $z_{850} < 24$.   We measure structural and size parameters
using these stellar mass maps, as well
as on ACS $BViz$ band imaging. This includes investigating the 
structural $CAS-$Gini-$M_{20}$ parameters
and half-light radius ($R_{\rm e}$) for each galaxy.  We further identify and
examine unusual galaxy types with this method, including compact and
peculiar ellipticals, and peculiar galaxies in some mode of
formation.  We compare structural parameters and half-light radii in the 
ACS $z_{850}$-band and stellar mass maps, finding no systematic bias 
introduced by measuring galaxy sizes in $z_{850}$.     We 
furthermore investigate relations between structural
parameters in the ACS $BViz$ bands and stellar mass maps, and
compare our result to previous morphological studies. Combinations of 
various parameters in stellar mass generally reveal clear separations 
between early and late type morphologies, but cannot easily 
distinguish between star formation and dynamically disturbed systems.  
We also show that while ellipticals and early-type spirals have
fairly constant CAS values at $z < 1$ we find a tendency for late-type
spiral and peculiar morphological types to have a higher $A(M_{*})$ at higher
redshift.   We argue that this, and the large fraction of peculiars that
appear spiral-like in stellar mass maps, are possible evidence for either
an active bulge formation  in some late-type disks at $z < 1$ or the presence of
minor merger events.

\end{abstract}


\section{Introduction}

One of the most intriguing questions in modern astronomy is how the universe
assembled into the structures we see today. A major 
part of this question is understanding how galaxies formed over
cosmic time.  A popular and  rapidly developing method of tracing the 
evolution of galaxies is through examining galaxy morphologies and
structures over a range of redshifts (e.g. Conselice 2003; Conselice, 
Rajgor \& Myers 2008; Conselice et al. 2009; Lotz \etal 2008; Cassata
et al. 2010; Ricciardelli
et al. 2010; Weinzirl et al. 2011). While understanding how the morphologies of 
galaxies evolve, and how matter in galaxies is structured, is fundamental, 
structural studies of galaxies have thus far focused on measuring properties 
in one or more photometric bands, 
and using this as a tracer of
evolution.  These types of structural analyses have always been measured in
terms of relative luminosities, such that brighter parts of galaxies
contribute more towards their structure and morphology.  To understand 
galaxies more fully however, it is important to study the evolution of 
galaxy structure in terms of their stellar mass distribution, as this better
reflects the underlying distribution of stars in a galaxy.

Morphological studies have evolved from initial attempts to describe the 
range of galaxy forms during the early-mid 20th century, towards modern 
efforts of linking the spatial distribution of a galaxy's stars to its 
formation history. In the move away from visual classifications, the goal 
has been to quantify galaxy structure/morphology in a way such that 
structures can be measured automatically and in a reliable 
way. There are two broad approaches for automated classification of
galaxies - the parametric and non-parametric methods.

In the parametric approach, the light profiles of galaxies, as seen in the
plane of the sky, are compared with predetermined analytic functions. For
example, single Sersic profiles are fit to an entire galaxy's light
profile.  Similarly, 
bulge-to-disc ($B/D$) light ratios are computed by fitting
two component profiles to a galaxy. These 
methods are however  unable to parameterise directly any star formation or
merging activity which produces random or asymmetric structures within
galaxies.
The ``non-parametric'' structural approaches have no such presumed 
light profiles implicit in their analyses (e.g., Shade et al. 1995; Abraham et al.
1996; Conselice 1997, 2003; Lotz et al. 2004). This is a more 
natural approach towards measuring galaxy structures, as the majority of 
galaxies in the distant Universe have irregular structures that are not
well fit by parameterised forms
(e.g. Conselice \etal 2005; Conselice, Rajgor \& Myers 2008). Perhaps the 
most successful and straightforward of the non-parametric
systems is the $CAS$ method, which uses a combination of concentration ($C$),
asymmetry ($A$) and clumpiness ($S$) values to separate galaxy types 
(Conselice \etal 2000; Bershady \etal 2000;  Conselice 2003).

Besides overall structure, the measurement of the size evolution of galaxies, as
measured through half-light radii,
also has important implications for their formation histories. For example, 
many studies such as Trujillo \etal (2007) and Buitrago et al. (2008) have
 measured galaxy sizes for systems at $z > 1$ and have found
a population of compact spheroid galaxies with number densities two orders of 
magnitude higher than what we find in the local universe. The absence of a significant 
number of these small sized (as measured by half-light radii), high mass galaxies 
in the local Universe suggests that this population has evolved and has perhaps merged with 
other galaxies. However, these studies are measured based on the light originating from 
these galaxies, and it is not clear whether sizes would change when measured 
using the distribution of stellar mass rather than light.

All of these methods for measuring the resolved structures 
of high redshift galaxies depend on measurements made in one or more 
photometric bands. As such, quantitative structures are  
influenced by a combination of effects, including: regions of enhanced 
star formation,  irregular dust distributions, differing ages and 
metallicities of stellar populations and minor/major mergers.
The morphology and structure of
a galaxy also depends strongly upon the rest-frame wavelength probed
(e.g., Windhorst et al. 2002; Taylor-Mager et al. 2007).
Young stars, such as OB stars can dominate the appearances of galaxies
at blue wavelengths, thereby giving a biased view of the underlying
mass within the galaxy.  Longer wavelength observations improve the
situation, but at every wavelength, the light emitted is from a mixture
of stars at a variety of ages and the dust content.    
 More fundamental is the structure  
of the stellar mass distribution with a galaxy, as it is more of a direct tracer
of the underlying potential. 
Although galaxy structure has been measured on rest-frame near-infrared and $I$-band 
imaging, which traces stellar mass to first order, we directly investigate the
stellar mass images within this paper. 

We use the method outlined in Lanyon-Foster, Conselice
\& Merrifield (2007; LCM07) to
reconstruct stellar mass maps of galaxies within the GOODS fields at $z < 1$. 
We then use this to investigate the evolution of the distribution of
galaxy stellar mass during the last half of the universe's history.
We use these stellar mass maps to directly measure 
$CAS$ parameters, and the sizes of these galaxies in stellar mass
over this epoch. We test the assumptions that the $CAS$
parameters and sizes 
can be reliably measured in optical light by comparing the parameters 
for these galaxies measured in  $the 
B_{435}$, $V_{606}$, $i_{775}$, and $z_{850}$ bands  and in stellar mass. 
We finally investigate how these various wavelengths and stellar mass
maps can be used to classify galaxies by their formation modes, revealing 
how these systems are assembling.



This paper is organised as follows: in \S~\ref{sec:data} and
\S~\ref{sec:method} we describe the data, sample selection and method,
including explanations of the  K-correction code we use and the $CAS$ 
analysis. In \S~\ref{sec:results} we present our results, discussing the 
stellar mass maps themselves, the comparison between galaxy size and the 
$CAS$ parameters in $z_{850}$ and mass. We then explore the 
relations between the structural parameters in $B_{435}$, $V_{606}$, 
$i_{775}$, $z_{850}$ and stellar mass. Finally, in \S~\ref{sec:conclusions}, 
we discuss our conclusions and comment on future applications.  Throughout 
we assume a standard cosmology of {\it H}$_{0} = 70$ km s$^{-1}$ Mpc$^{-1}$, 
and $\Omega_{\rm m} = 1 - \Omega_{\Lambda}$ = 0.3.

\section{Data and Sample}\label{sec:data}

\subsection{Data}

The primary source of our data consists of HST/ACS imaging from the GOODS ACS 
imaging Treasury Program\footnote{http://archive.stsci.edu/prepds/goods/}. 
The observations consist  of imaging in the $B_{435}$ (F435W), $V_{606}$ 
(F606W), $i_{775}$ (F775W) and $z_{850}$ (F850LP) pass-bands, covering the 
Hubble Deep Field-North (HDF-N) area.    The central wavelengths of these
filters, and their full-width at half-maximum,
are: F435W (4297, 1038 \AA), F606W (5907, 2342 \AA), 
F775W (7764, 1528 \AA), F850L (9445, 1229 \AA).
 The images have been reduced (using the ACS CALACS pipeline),
calibrated, stacked and mosaiced by the GOODS team (Giavalisco \etal
2004). The field is provided  in many individual image sections, 
with the HDF-N data divided into 17 sections, each of
$8192\times8192$ pixels.  The total area of the GOODS survey is roughly
315 arcmin$^{2}$, and we utilise imaging which was drizzled  with
a pixel scale of 0.03 arcsec pixel$^{-1}$, giving an effective PSF
size of $\sim 0.1$\arcsec in the $z-$band.

The sample consists of galaxies selected in the GOODS-N field with $z \leq
1$ and a F814W magnitude of $< 24$. The sample was also
restricted to those galaxies with 
data in all  four ACS photometric bands such that we can 
subsequently obtain the most 
accurate K-corrections and stellar masses possible. The final sample 
consists of 560 objects, which were individually cut out of the GOODS-N 
ACS imaging, and handled as separate entities. This is described in detail 
in the  following section.

Photometric redshifts are available for the whole sample (Mobasher \etal
2004), and spectroscopic redshifts are available for 404 out of 560 galaxies,
as found by the Team Keck Redshift Survey (TKRS; Wirth \etal, 2004). When
spectroscopic redshifts are available they are used, else the photometric
values are substituted.

\section{Method}\label{sec:method}

\subsection{Stellar Mass Maps}

To create stellar mass maps of our galaxies, each pixel in each
galaxy image is treated individually
throughout the analysis. We convert fluxes from counts per pixel to apparent
magnitudes per square arc-second for each pixel in every image in all four
photometric bands, as well as calculating their associated errors. 
Pixels with negative  fluxes
were assigned values four orders of magnitude smaller than the typical  flux
so that even pixels with low signal-to-noise values can be mapped. This 
however means that our resulting values for the CAS parameters, especially
asymmetry which uses the background light to do a correction, are potentially
different than when measured using optical light.

Because our galaxies are at a variety of redshifts, we have to carry out
fitting to each SED for each pixel to calculate the stellar mass within
each pixel.  
K-corrections and stellar masses are calculated for each pixel of each image
using the ``K-Correct'' code of Blanton \& Roweis (2007).  This is
similar to previous methods outlined by Lanyon-Foster et al. (2007),
Bothun (1986), Abraham et al. (1999), and Welikala et al. (2008, 2011).

The code, K-Correct\footnote{distributed at
http://cosmo.nyu.edu/blanton/kcorrect/}
naturally handles large datasets, and through its interpretation of the data
in terms of physical stellar population models, outputs results in terms of
stellar mass and star formation histories (SFHs) of each galaxy. The code
allows the input of GOODS data and outputs a stellar mass for each object. The
code can also be modified so as to treat each pixel as a separate input object,
which we do in this work.

The K-correction ($K_{QR}(z)$) between bandpass $R$, used to observe a galaxy
with apparent magnitude ($m_{R}$), at redshift $z$, and the desired 
bandpass $Q$ is defined as (e.g., Oke \& Sandage 1968):

\begin{equation} \label{eq:kcdef}
m_R = M_Q + DM(z) + K_{QR}(z) - 5log(h)
\end{equation}

\noindent where	

\begin{equation} \label{eq:dmod}
DM(z) = 25 + 5 log \left[ \frac{d_{L}}{h^{-1} Mpc} \right]
\end{equation}

\noindent is the bolometric distance modulus calculated from the luminosity
distance, $d_L$, $M_Q$ is the absolute magnitude and $h$ = H$_{0}$/100
km\,s$^{-1}$\,Mpc$^{-1}$. 

The K-correct software contains model templates in electronic form
and an implementation of the method to fit data to models. The code uses
the stellar population synthesis models of Bruzual \& Charlot (2003) and
contains training sets of data from GALEX, SDSS, 2MASS, DEEP2 and GOODS.
The code finds the nonnegative linear combination of
$N$ template star formation histories that best match the observations
using a minimum $\chi^2$ comparison. With the entire set of galaxy
observations available, K-correct also fits for the $N$ template SFHs using
a nonnegative matrix factorisation algorithm.   K-correct
naturally handles data uncertainties, missing data, 
and deals with the complications of observing galaxy spectra
photometrically, using broadband filters of galaxies at varying redshifts.

We use a set of 485 spectral templates to fit to our galaxy pixel
SEDs using Bruzual \& Charlot (2003) models with 
the Chabrier (2003) IMF and Padova (1994) isochrones. All of the six metallicities 
available (mass fractions of elements heavier than He of $Z = 0.0001, 0.0004, 0.004, 0.008,
0.02$ and $0.05$) are used.

Some of the known areas of uncertainty in the stellar population models are in
the UV and IR regions. In the UV light from young or intermediate aged
stellar populations can dominate the flux at
$\sim 1500 \AA$.
In the near-IR, thermally pulsating asymptotic giant branch (TP-AGB) stars
dominate the flux in some intermediate age populations (Maraston, 2005). We
discuss this issue in terms of stellar masses in Conselice et al. (2007) who
show that the effects of TP-AGB stars do not kick in until at
redshifts higher than the scope of this study ($z > 2$), and thus are not a
concern in the present work.

Results are outputted from K-correct for each
input pixel, giving rest-frame fluxes at $NUV$, $U$, $B$, $V$, $R$, $I$ 
magnitudes
plus their associated errors, mass-to-light ratios in all of these bands and
finally the stellar mass in that pixel.  We then reconstruct the image of
the galaxy in stellar mass based on these calculations across the galaxy.  
We use a signal-to-noise limit of S/N $= 3$ per pixel in stellar mass for all
optical bands, for a reliable calculation.  We also investigate how the
results would change if pixels are summed together, finding similar results
with the exception of the asymmetry which tends to decrease at lower 
resolutions.

\begin{figure}
\includegraphics[angle=0, width=85mm, height=85mm]{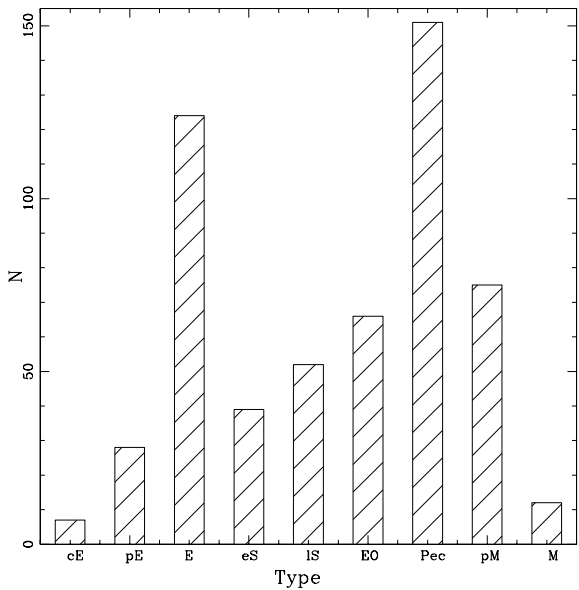}
\caption{The number of galaxies in each of our sample classifications
as described in \S 3.2.   Galaxies were visually classified by
  inspection of both $z_{850}$ images and segmentation maps.}
\label{fig:class}
\end{figure}

\subsection{Visual Classifications}

We have classified our entire sample visually, using the $z_{850}$ galaxy
images and the object segmentation maps. We have derived the classification
system by applying the  most appropriate criteria to this particular
dataset. The classification scheme is divided into nine categories, ranging
from  compact objects (spherical with little or no apparent envelopes) to
obviously merging systems (with evidence of tidal streams and  other merger
attributes). Six objects in the catalogue were found to be unresolved and
these were removed from the sample.  We give a description of these types
and how they were selected below.  We often refer to these galaxy types
throughout this paper, and Appendix B gives a description of these
types in terms of our measured indices.  Figures 2-10 show examples of
these galaxy types.

\begin{figure}
\includegraphics[angle=0, width=85mm, height=130mm]{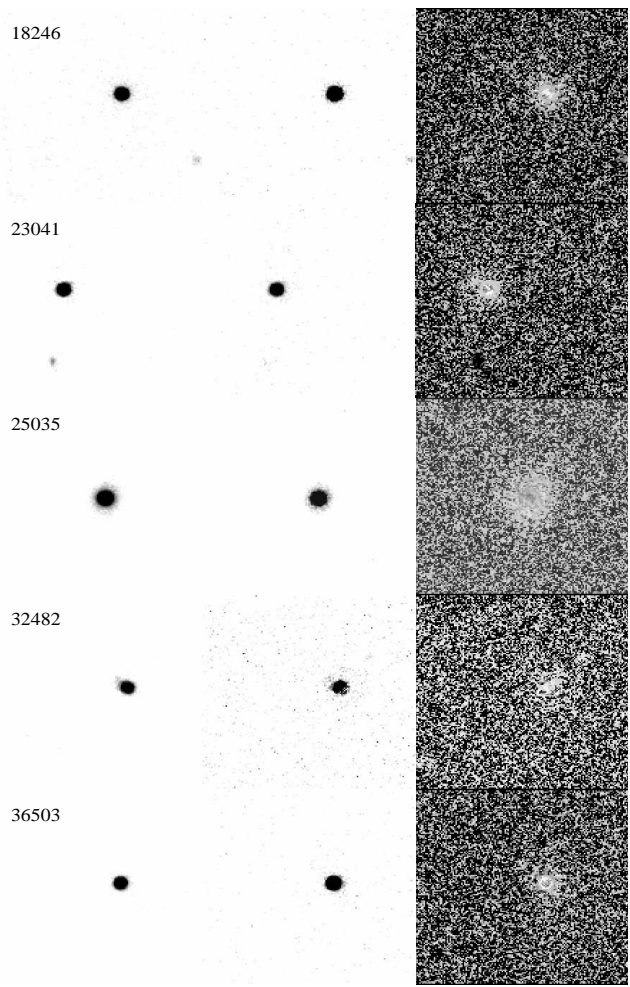}
\caption{Visual displays of the light (in $z_{850}$, left), stellar mass (middle) and the distribution of $M/L$ ratios (right) for the compact elliptical (cE) galaxies.  The $M/L$ map is scaled such that white corresponds to higher $M/L$ values, and black to lower $M/L$ values. This convention is used in the following  Figures ~\ref{fig:massmap_pE} to ~\ref{fig:massmap_M}.}
\label{fig:massmap_cE}
\end{figure}

\begin{figure}
\includegraphics[angle=0, width=85mm, height=130mm]{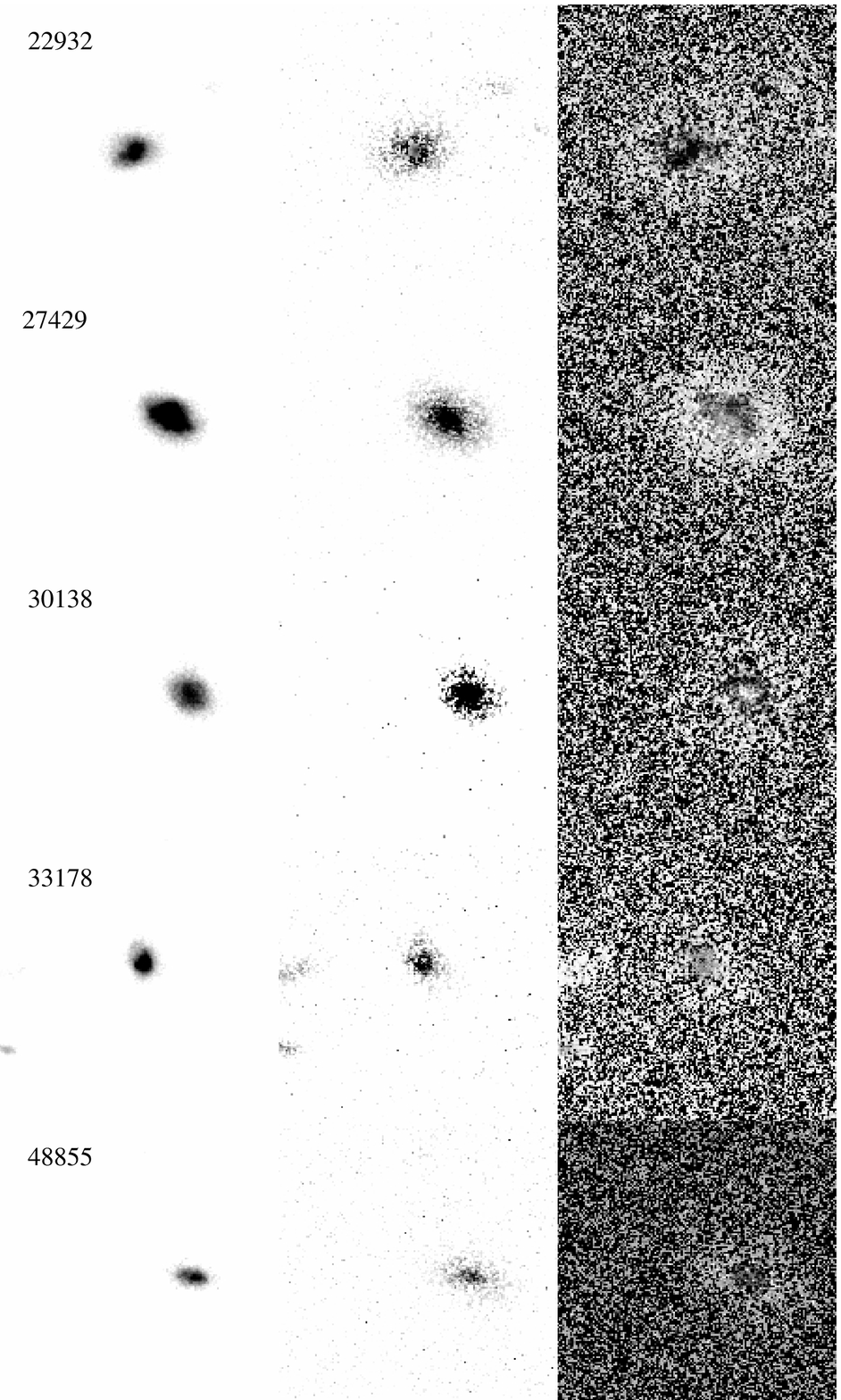}
\caption{Comparing the light (in $z_{850}$), stellar mass maps, and the 
distribution of $M/L$ ratios for examples of peculiar elliptical (pE) 
galaxies. The scale and convention is the same as that used in 
Fig. ~\ref{fig:massmap_cE}.}
\label{fig:massmap_pE}
\end{figure}

\subsubsection{Compact Ellipticals}

The {\em compact ellipticals}, denoted as ``cE'', were identified as compact 
objects that varied smoothly across their radii when viewed morphologically. 
These objects had little or no sign of an envelope usually 
associated with early-type galaxies.
The compact Es are generally found at lower redshifts, at $z < 0.4$
with a few between $0.6 < z < 0.8$.  The distinguishing feature
of these compact Es is their small sizes, with half-light radii
of $< 1$ kpc.

\subsubsection{Ellipticals}

The elliptical galaxies are amongst the most popular and important
galaxy type in our study, as they are typically the most massive galaxies
in the local universe.   One reason for this is that elliptical
galaxies, and massive galaxies in general, are the test-bed
for understanding theories of galaxy formation which often
have strong predictions for how the most massive galaxies
should form (e.g., Bertone \& Conselice 2009). The ellipticals we see at high
redshift are likely the progenitors of the massive
ellipticals found today.

Galaxies with the familiar features of nearby {\em elliptical galaxies} were 
classed as such, and given the label ``E".    28 E galaxies were identified, 
which at first glance have the appearance of early
types, but on closer inspection and by varying the image contrast had some
peculiar features, such as multiple nuclei or minor disturbances in the
internal structure. These objects were classified as 
{\em peculiar ellipticals}, and are labelled as  pE (Conselice \etal  2007).

\subsubsection{Spirals/Disks}

Objects with a disc-plus-bulge structure
were classified as {\em early-type spirals} (``eS") if the bulge appears
larger than the disc component, and {\em late-type spirals} (``lS") if the 
disc is larger than the central bulge. Edge-on disc galaxies are denoted  
by ``EO".

The spiral galaxies are amongst the most interesting for
this study given that they often contain two major stellar
population types segregated spatially. Traditionally 
this is seen as an older stellar population making up
the bulge, and the spiral arms consisting of
younger stellar populations.  

\subsubsection{Peculiars and Mergers}

Objects that did not fit easily into any of the previously defined categories,
but whose segmentation map showed them to be singular, or unconnected to any
apparently nearby object in the image, were classified as {\em peculiar},
``Pec". These are galaxies that could possibly have merged in the recent past.

Galaxies whose morphology also ruled them out from any previously defined
class but whose segmentation map showed them to be connected to, or associated
with, at least one other object on the image, but without obvious merger
signatures, were classified as {\em possible mergers} (``pM"). Galaxies 
with the pM
requirements and which showed obvious signs of merging, such as tidal tails,
were classified as {\em mergers} (``M").

This information is summarised for quick reference in Table
\ref{tab_classfcn}. The sample was independently classified four times. 
The mode of the results
were taken to  be the true classification and where an indecisive split was
produced the lowest numerically valued type was chosen. Figure
~\ref{fig:class} shows a histogram of the sample classifications.

We note that there are a significant number of peculiar galaxies in our 
sample (e.g., Figure~1), much higher than what has been found at lower 
redshifts in previous work investigating galaxy morphology (e.g., Conselice 
et al. 2005). 
These are systems that at lower resolution would be classified as disk-like 
galaxies in many cases. They are not necessarily merging systems, but are 
likely normal galaxies in some type of formation, the modes of which are one
of the features we investigate later in this paper.

We show examples of each of the above types  in Figures 2-10, with
five examples shown for each. These figures show the image of the galaxy
in the $z_{850}$-band, the stellar mass maps, and the mass to light ratio for each
galaxy.

\begin{table}
\centering
\begin{tabular}{|l|l|}
\hline
Type   &   Description  \\
\hline
E       &   Elliptical   \\
cE      &   Compact E     \\
pE      &   Peculiar E   \\
eS      &   Early-Type Spiral \\
lS      &   Late-Type Spiral  \\
EO      &   Edge-on disc \\
Pec     &   Peculiar    \\
pM      &   Possible merger  \\
M       &   Obvious merger  \\

\hline

\end{tabular}
\caption{Descriptions of the classification scheme.}
\label{tab_classfcn}
\end{table}

\begin{figure}
\includegraphics[angle=0, width=85mm, height=130mm]{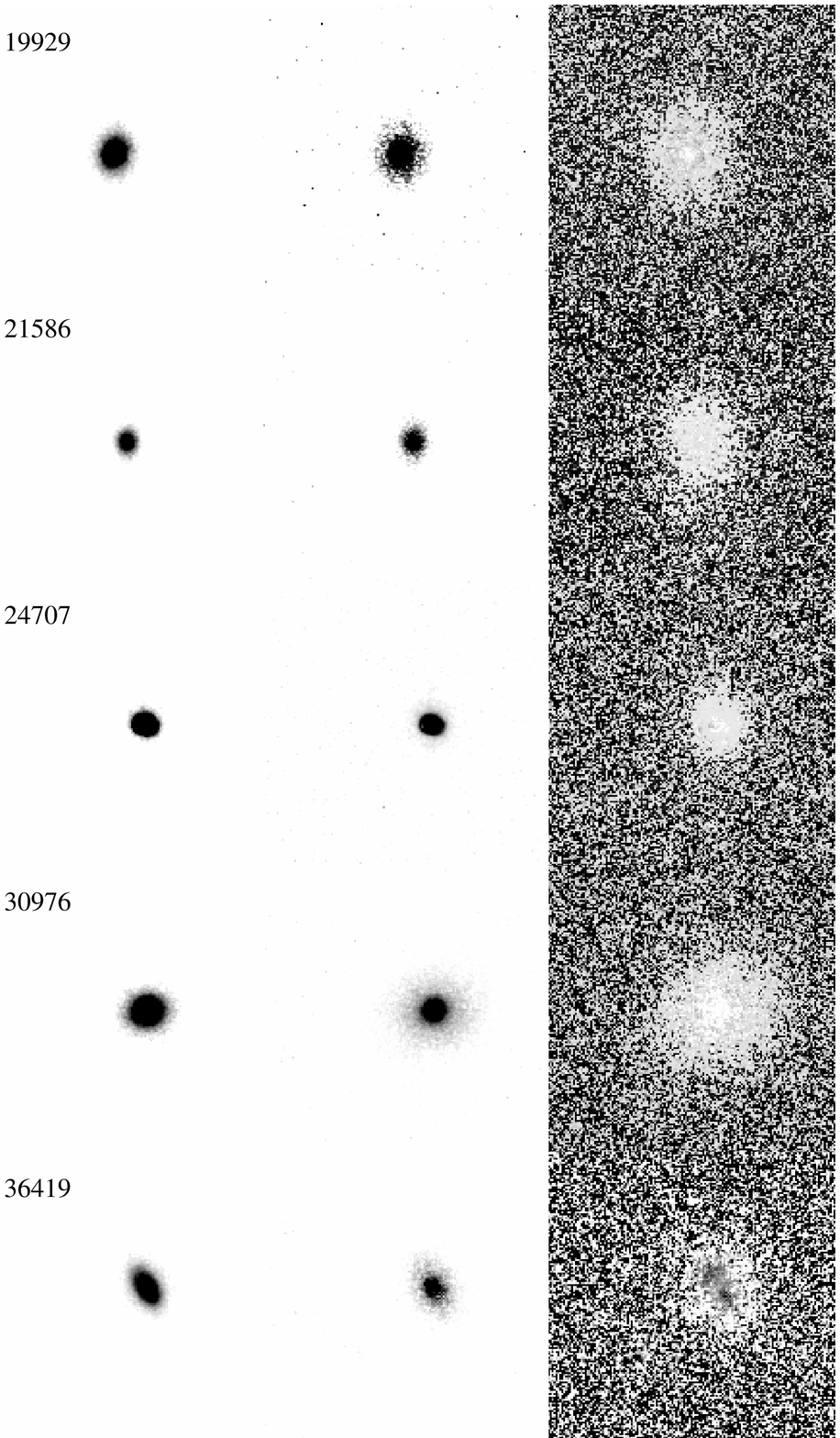}
\caption{Comparing light, stellar mass maps and $M/L$ for the elliptical (E) galaxies. The scale and convention is the same as that used in Fig. ~\ref{fig:massmap_cE}.}
\label{fig:massmap_E}
\end{figure}

\begin{figure}
\includegraphics[angle=0, width=85mm, height=130mm]{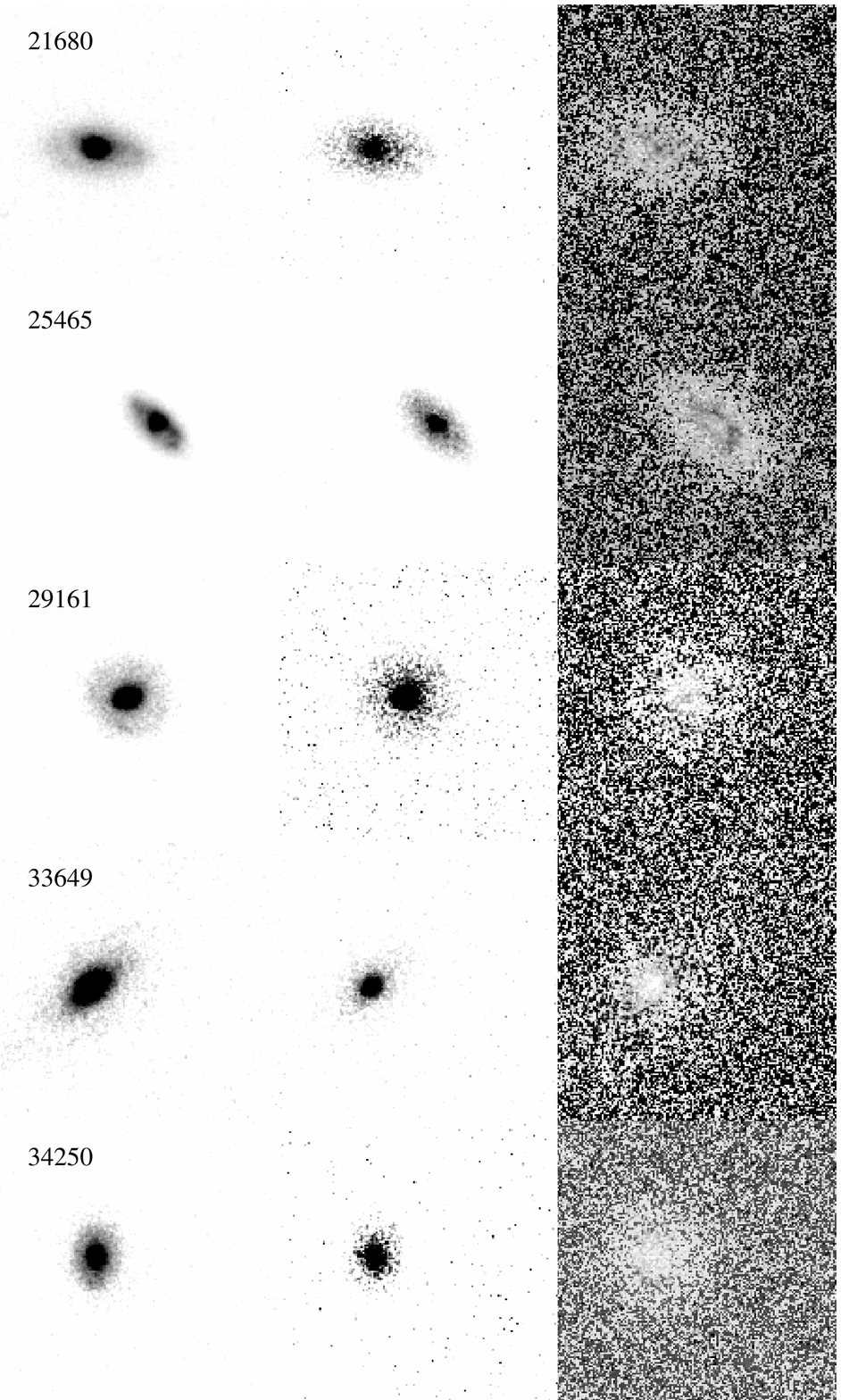}
\caption{Comparing light, stellar mass maps and $M/L$ for the early-type spiral galaxies (eS). The scale and convention is the same as that used in Fig. ~\ref{fig:massmap_cE}.}
\label{fig:massmap_eS}
\end{figure}

\begin{figure}
\includegraphics[angle=0, width=85mm, height=130mm]{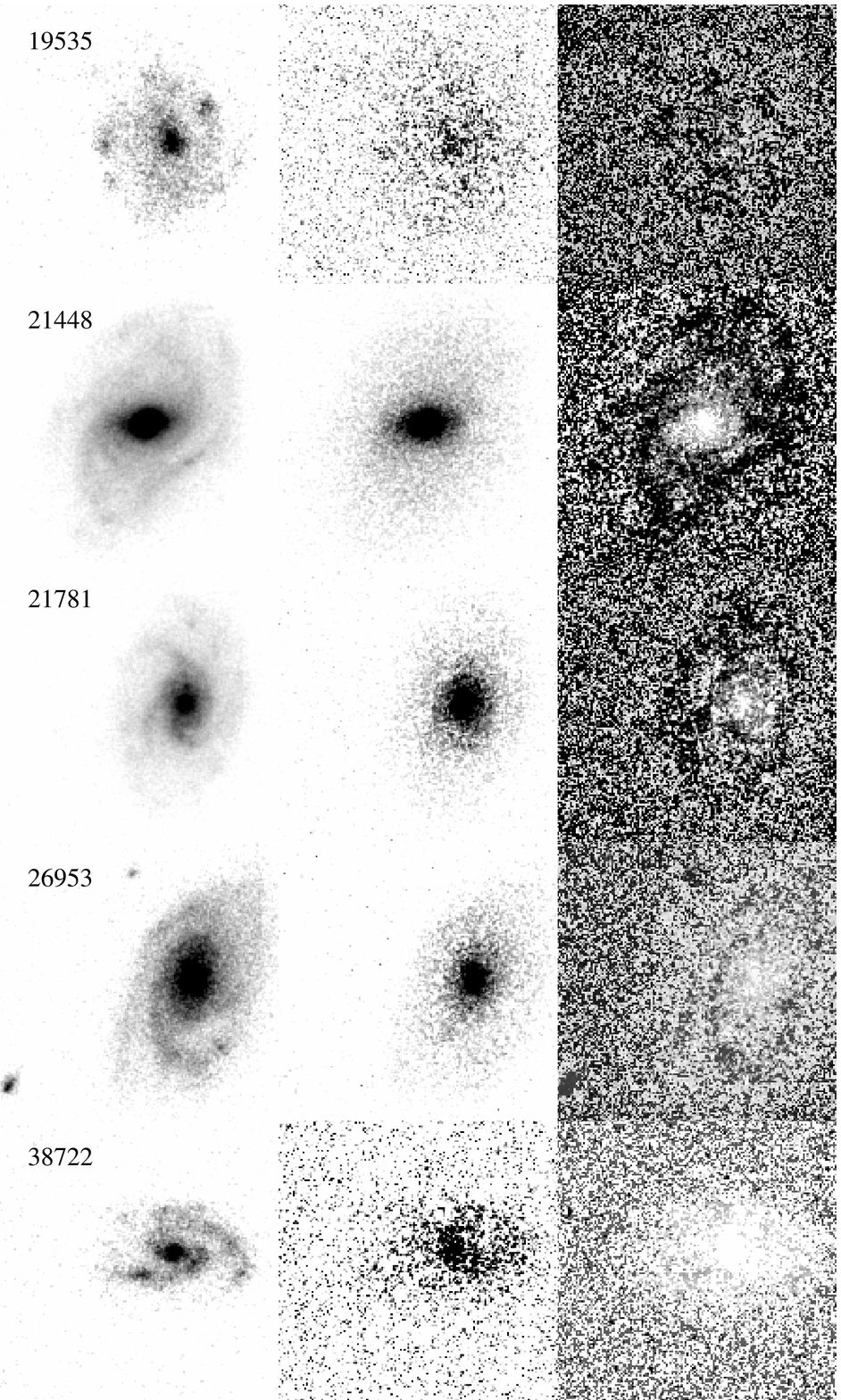}
\caption{Comparing light, stellar mass maps and $M/L$ for the late-type spiral galaxies (lS). The scale and convention is the same as that used in Fig. ~\ref{fig:massmap_cE}.}
\label{fig:massmap_lS}
\end{figure}

\begin{figure}
\includegraphics[angle=0, width=85mm, height=130mm]{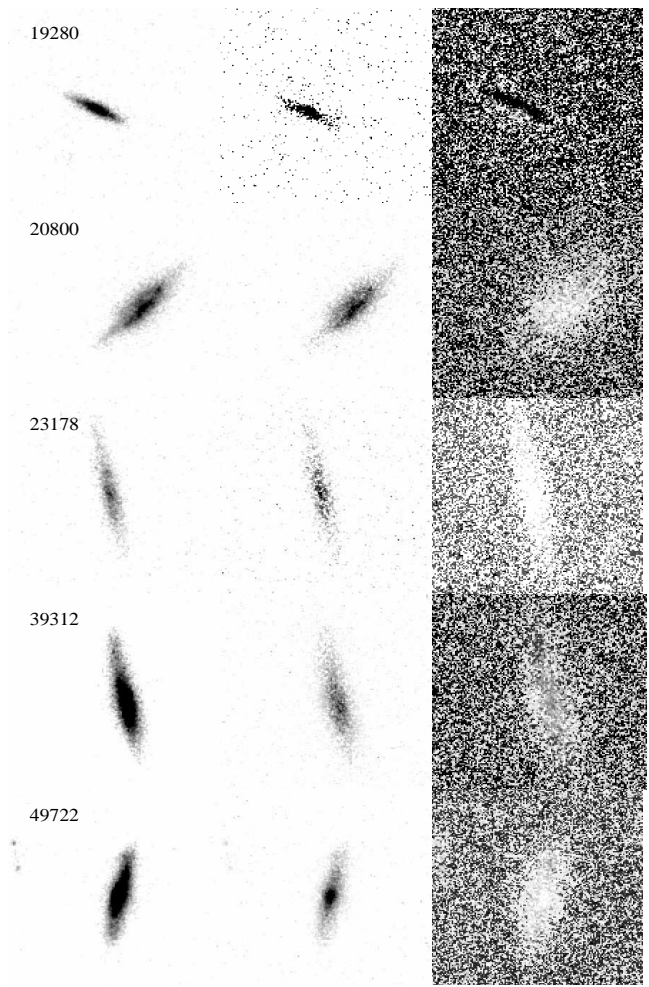}
\caption{Comparing light, stellar mass maps and $M/L$ for the edge-on disk galaxy (EO) galaxies. The scale and convention is the same as that used in Fig. ~\ref{fig:massmap_cE}.}
\label{fig:massmap_EO}
\end{figure}

\begin{figure}
\includegraphics[angle=0, width=85mm, height=130mm]{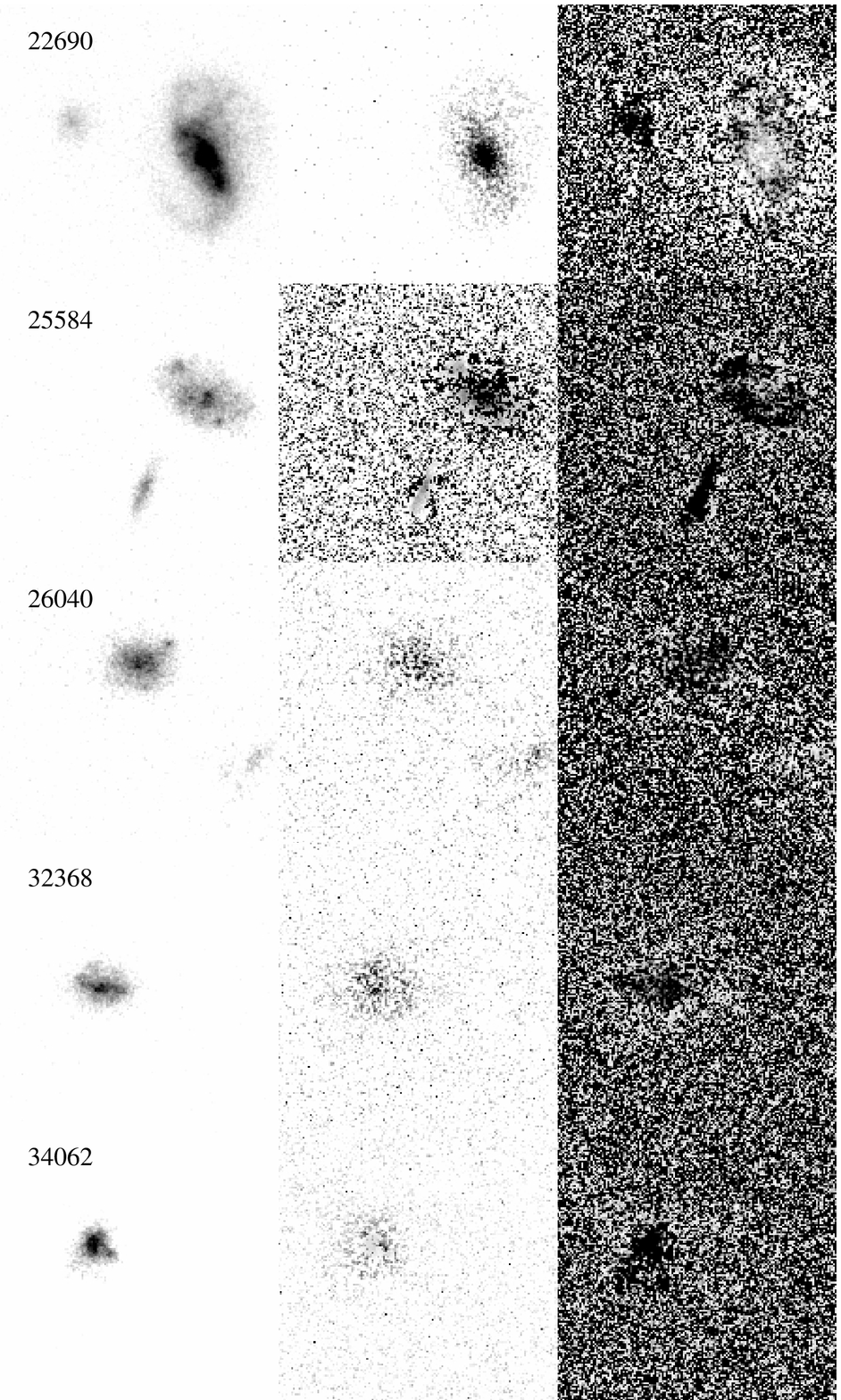}
\caption{Comparing light, stellar mass maps and $M/L$ for the peculiar (Pec) galaxies. The scale and convention is the same as that used in Fig. ~\ref{fig:massmap_cE}.}
\label{fig:massmap_Pec}
\end{figure}

\begin{figure}
\includegraphics[angle=0, width=85mm, height=130mm]{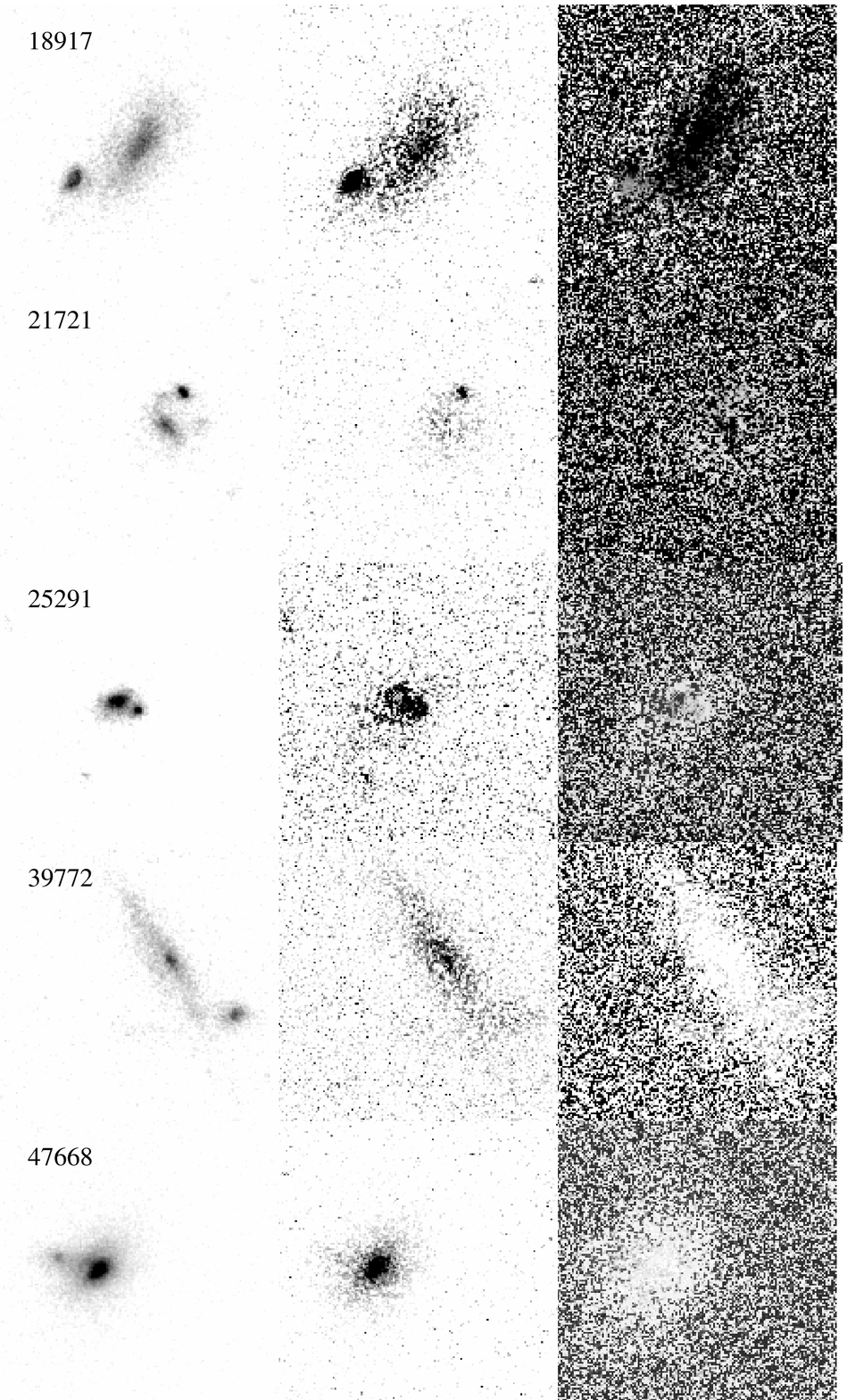}
\caption{Comparing light, stellar mass maps and $M/L$ for the possible-merging (pM) galaxies. The scale and convention is the same as that used in Fig. ~\ref{fig:massmap_cE}.}
\label{fig:massmap_pM}
\end{figure}

\begin{figure}
\includegraphics[angle=0, width=85mm, height=130mm]{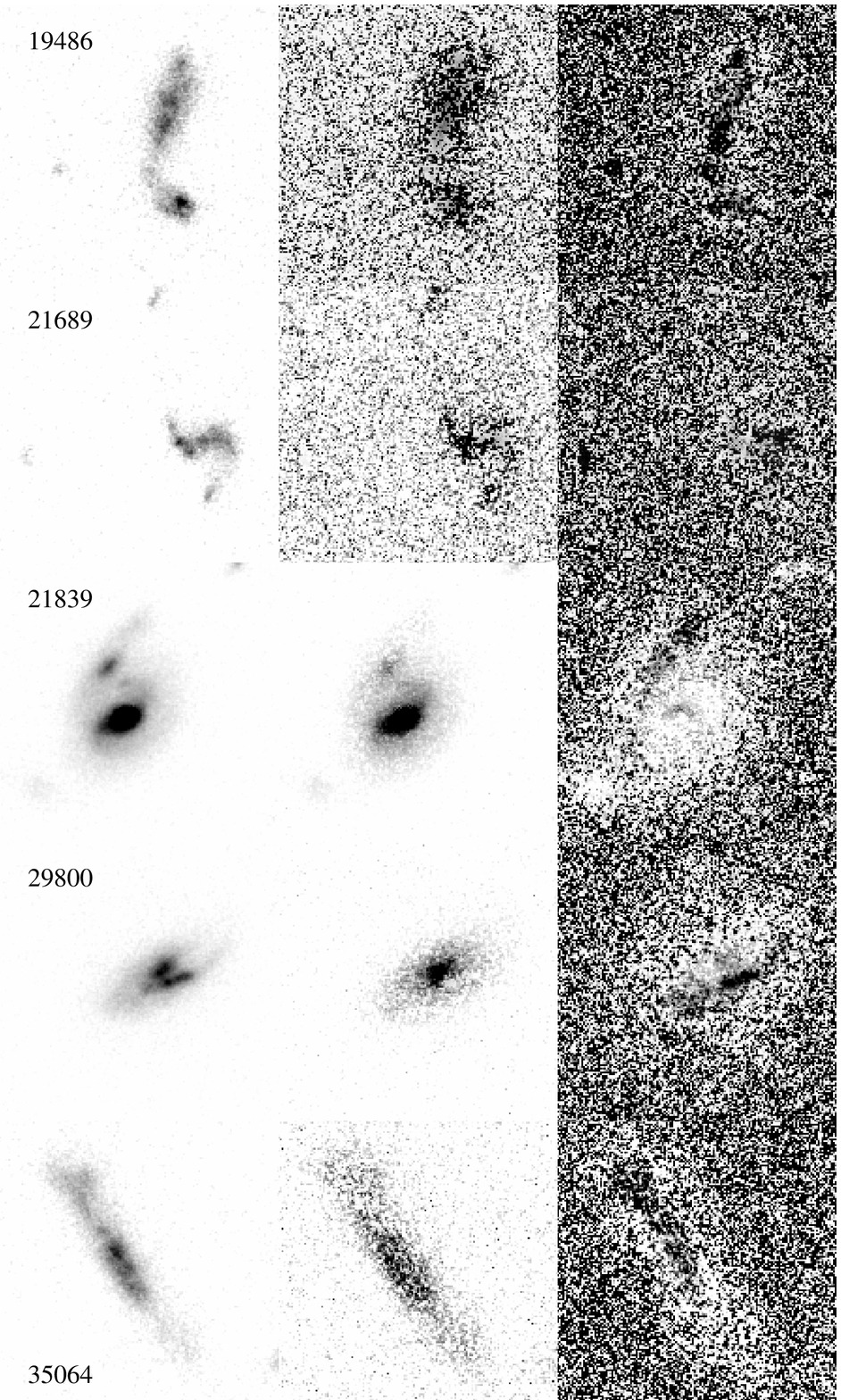}
\caption{Comparing light, stellar mass maps and $M/L$ for the merging (M) galaxies. The scale and convention is the same as that used in Fig. ~\ref{fig:massmap_cE}.}
\label{fig:massmap_M}
\end{figure}

\subsection{CAS Analysis}\label{sec:analysis}

We use the concentration, asymmetry, clumpiness ($CAS$) parameters to
quantitatively measure the structures of our sample, in all available bands,
$BViz$, and on the stellar mass maps.  We also measure the Gini and M$_{20}$ 
parameters, forming an extensive non-parametric method for measuring the
structures and morphologies of galaxies in resolved CCD images (e.g.,
Conselice et al. 2000a; Bershady et al. 2000; Conselice et al. 2002; Lotz
et al. 2004;
Conselice 2003; Lotz et al. 2008). The premise for using these parameters is
to tap into the light distributions of galaxies, which reveal their past and
present formation modes (Conselice 2003). The regions into which the
traditional Hubble types fall in CAS parameter space is well understand
from local galaxy comparisons. For
example, selecting objects with $A > 0.35$ finds systems that are
highly disturbed, and nearly all are major galaxy mergers (e.g., Conselice et
al. 2000b; Conselice 2003; Hernandez-Toledo et al. 2005; Conselice 2006a).
A more detailed
analysis of this problem is provided in Appendix A for optical light.

\begin{figure*}
\includegraphics[angle=0, width=148mm, height=138mm]{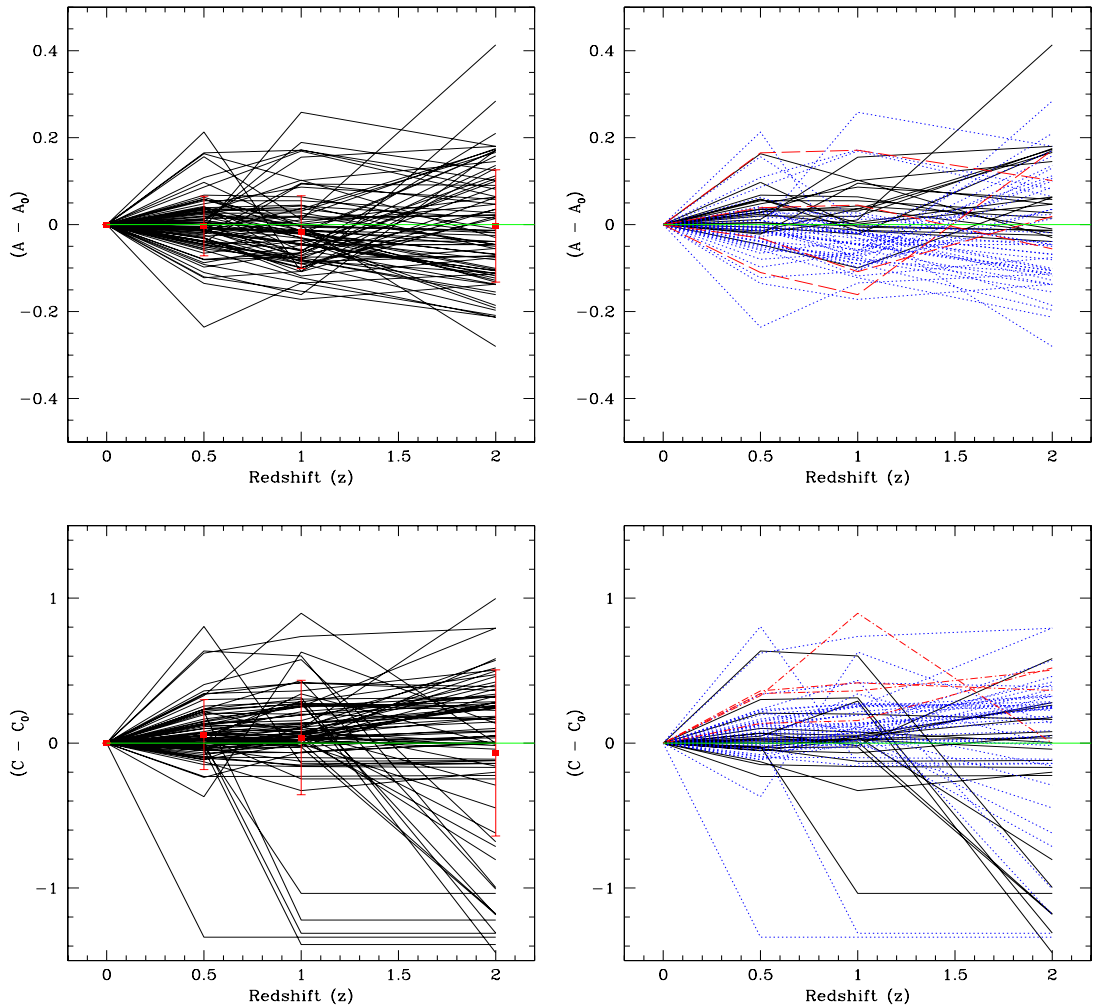}
\caption{The change in the asymmetries and concentrations measured on our
stellar mass maps as a function of
redshift due simply to distance effects.  Show at the top is the change in the asymmetry
parameters for a mixture of nearby galaxies of various types: ellipticals, spirals and
irregulars, while the bottom two panels are for the concentration index.   The left hand
side for both shows the change for the entire sample we simulation with the red dots and
errorbars showing the average change and 1 $\sigma$ variation of that change.  The right
hand side shows these simulations divided up between ellipticals (black solid line), 
spiral galaxies (blue dotted line) and irregulars (red dashed line.)}
\label{fig:sim}
\end{figure*}

\begin{figure}
\includegraphics[angle=0, width=90mm, height=90mm]{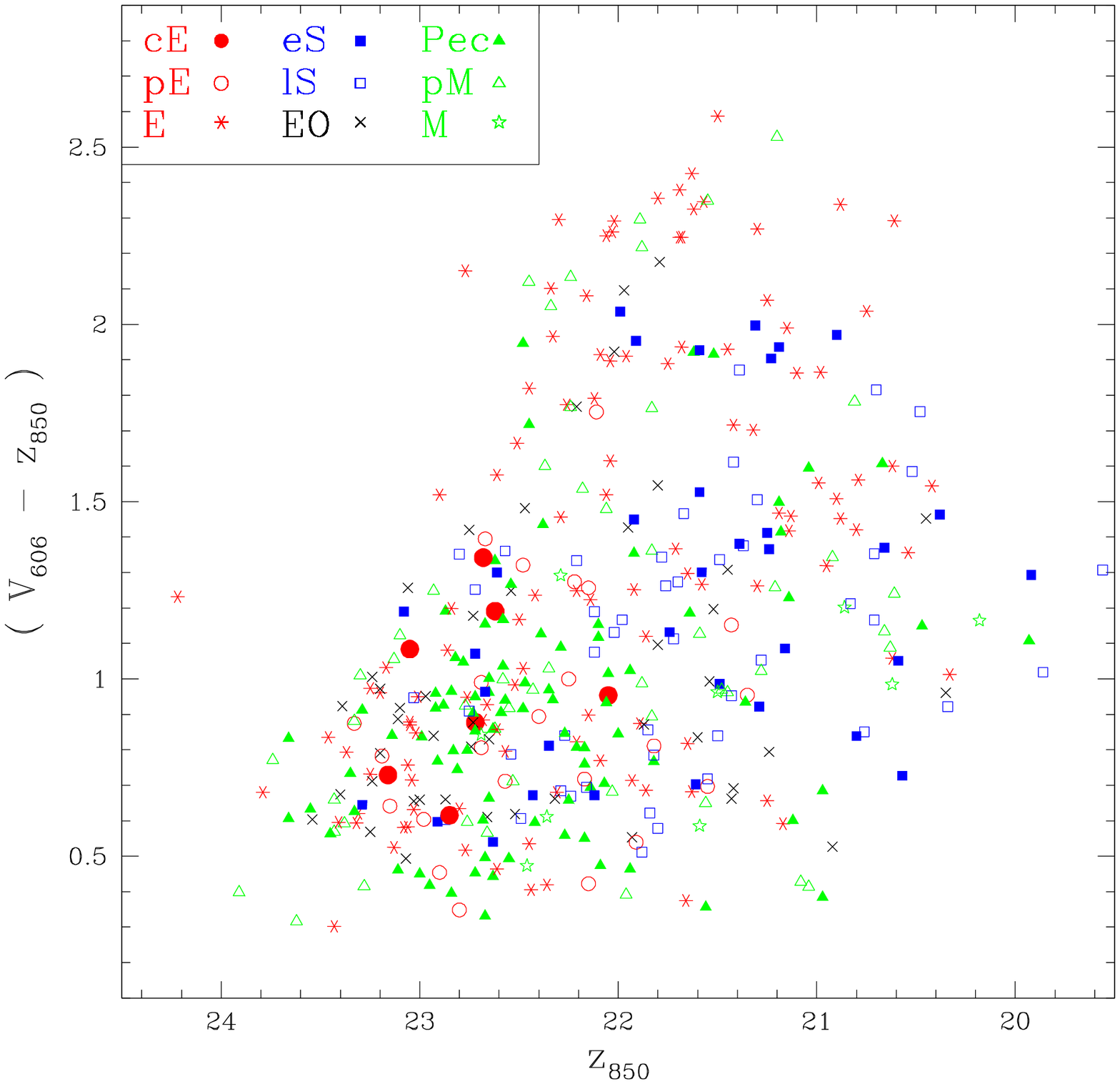}
\caption{The colour-magnitude relation for our sample of galaxies.
Displayed is the ($V_{606} - z_{850}$) colour vs. the magnitude of
the galaxy as observed in the $z_{850}$ band.  Different galaxy
types are displayed, including: compact ellipticals (cE), peculiar
ellipticals (pE), ellipticals (E), early-type spirals (eS), late-type
spirals (lS), edge-on disk galaxies (EO), peculiar galaxies (Pec),
possible-merger galaxies (pM) and merging systems (M).  This key is used
  throughout the remainder of this paper.}
\label{fig:CMD}
\end{figure}


We measure the structural parameters for the GOODS sample using the method of
Conselice \etal (2008), with slight adjustments made for the stellar mass 
maps, to enable the code to handle the large values of stellar masses per 
pixel, rather than flux.  The radius of each individual galaxy within the 
postage stamp image/mass map is measured on the stellar mass map, and we 
define all our indices within the Petrosian radii (e.g., Petrosian 1976; 
Bershady \etal 2000; Conselice 2003).  the Petrosian radius has been 
found to be a better are more reliable radius (and reproducible) than the
isophotal radius (Petrosian 1976).  
The limits and relationship to other radii is described in detail in
terms of total light in a galaxy by Graham et al. (2005).

Circular apertures are used for measuring our 
Petrosian radii and quantitative parameter estimation. 
The Petrosian radius used to measure our parameters is defined by,

$$R_{\rm Petr} = 1.5 \times r(\eta = 0.2),$$

\noindent where $r(\eta = 0.2)$ is the radius where the surface brightness 
(or stellar mass per unit area) is
20 percent of the surface brightness (or stellar mass per unit area)
within that radius (Bershady et al. 2000).  Note that this is a distance
independent measurement, given that surface brightness dimming effects
both measurements of surface brightness in the same way.  

Accounting for background light and noise is extremely important when
measuring structural parameters, especially for faint galaxies, and this 
must also be dealt with for the stellar mass maps. The measured 
parameters in the $B_{435}$, $V_{606}$, $i_{775}$ and $z_{850}$ bands and 
stellar mass images are corrected as described
in Conselice \etal (2008), by considering a background area close to the
object, and the segmentation map of the object. Using a background close to
the galaxy itself, any problems introduced by objects imaged on a large mosaic,
with a non-uniform weight map, and the object itself being faint compared to
the background, are alleviated. We review below how the
$CAS$ parameters are measured. These are described in more detail
in Bershady et al. (2000), Conselice et al. (2000), Conselice (2003) and Lotz
et al. (2008).


\subsubsection{Asymmetry}

We measure the asymmetry of a galaxy by taking the original galaxy image (or
stellar mass map), rotating it by 180 degrees about its centre, and then 
subtracting the two images (Conselice 1997), with corrections for background 
and radius (see Conselice \etal 2000a for details). The centre of rotation 
is found using an iterative process, which locates the minimum asymmetry. 
A correction is made within the stellar mass maps for this parameter, 
as it was found that uncertainties in the mass-to-light ratio adversely 
affected the asymmetry measurement. We found that approximately 15 percent 
of the stellar mass in the asymmetry calculation is left
over from random fluctuations in $M/L$.  This was compensated for including
an addition correction for this uncertainty through a slight additional
subtraction to the asymmetry signal.  This was done by scaling the 
asymmetric residuals of the rotated and subtracted image by an increase of
15\%, that is only 85\% of the residual is considered.

We measure the background
light in the same way as for the light measures of asymmetry, although because
we are dealing with a conversion to stellar mass, which is not trivially
done for the background, we have some background values which are lower
than zero.  To deal with this, we placed all the negative pixels in the
stellar mass map background to zero.  

The equation for calculating asymmetry is:

\begin{equation}
A = {\rm min} \left(\frac{\Sigma|I_{0}-I_{180}|}{\Sigma|I_{0}|}\right) - {\rm
  min} \left(\frac{\Sigma|B_{0}-B_{180}|}{\Sigma|I_{0}|}\right)
\label{eq:asym}
\end{equation}

\noindent Where $I_{0}$ denotes the original image pixels, $I_{180}$ is the
image after rotating by 180\deg. The background subtraction is made using
light (or mass) from a blank sky area, $B_{0}$, and is minimised using the same process
as for the object itself. The higher the
values of $A$, the higher the degree of asymmetry the galaxy possesses, the
most extreme cases usually corresponding to merger candidates (Conselice 2003).

\subsubsection{Concentration}

The concentration parameter measures the the intensity of light (or stellar 
mass) contained within a pre-defined central region, compared to a larger 
region towards the edge of the visible galaxy. Concentration is most often 
defined as the ratio of the flux contained within circular radii possessing 
20 percent and 80 percent ($r_{20}$, $r_{80}$) of the total galaxy flux,

\begin{equation}
C = 5 \times {\rm log} \left(\frac{r_{80}}{r_{20}}\right).
\end{equation}

\noindent A higher value of $C$ corresponds to an object where more light is
contained within the central region. This measurement has been shown to
correlate with the halo, and total stellar masses, of nearby 
galaxies (e.g. Bershady \etal 2000; Conselice 2003).

\begin{figure}
\centering
\includegraphics[angle=0, width=90mm, height=90mm]{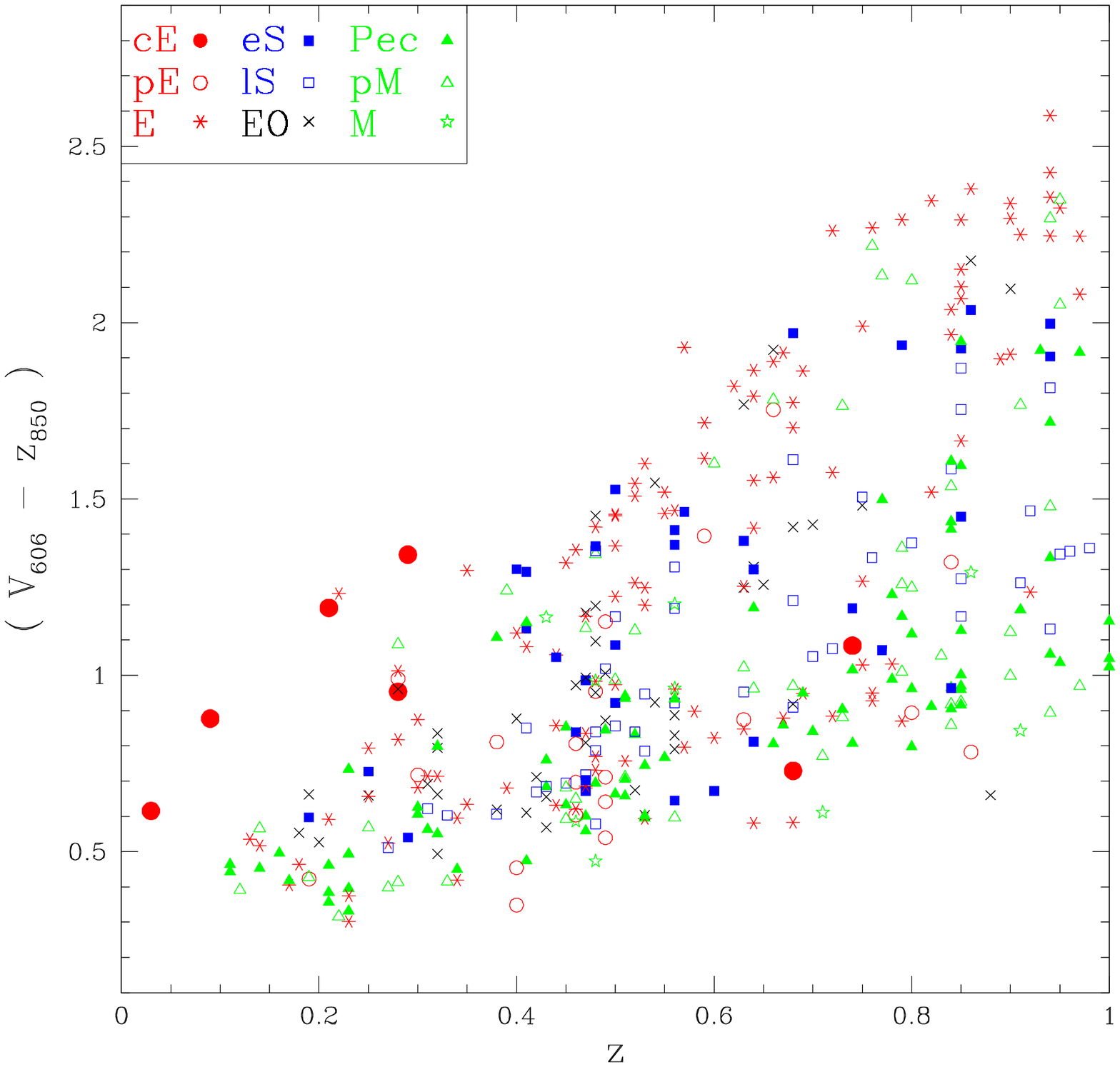}
\caption{Colour in observed ($V_{606} - z_{850}$) vs. Redshift ($z$) 
for our sample. The points displayed are the same as in Fig. ~\ref{fig:CMD}.}
\label{fig:colz}
\end{figure}

\subsubsection{Clumpiness}

Clumpiness ($S$) is related to asymmetry, in that it is used to measure the
amount of light (or mass) in a galaxy that exists in discrete, clumpy
distributions. In this definition, a smooth galaxy contains light at low
spatial frequencies, such as elliptical galaxies, whereas clumpy systems have
most of their light contained in high spatial frequencies. Star forming
galaxies tend to be clumpy in structure, with high $S$ values. We measure
clumpiness via:

\begin{equation} 
S = 10 \times \left[\left(\frac{\Sigma
(I_{x,y}-I^{\sigma}_{x,y})}{\Sigma I_{x,y} }\right) - \left(\frac{\Sigma
(B_{x,y}-B^{\sigma}_{x,y})}{\Sigma I_{x,y}}\right) \right],
\end{equation}

\noindent where, the original image $I_{x,y}$ is blurred to produce a
secondary image,  $I^{\sigma}_{x,y}$. The secondary image is subtracted from
the original image to create a residual map, showing only the high frequency
structures contained within the galaxy (Conselice 2003). The residuals are
quantified by normalising, using the total light in the original galaxy image,
and then subtracting the normalised residual sky. The smoothing kernel,
$\sigma$, used is determined from the radius of the galaxy, and has the value
$\sigma = 0.2 \cdot 1.5 \times r(\eta = 0.2)$ (Conselice 2003). The centres of
the galaxies (roughly the inner 1/10th the Petrosian radius) are removed 
during this procedure.  This parameter is furthermore the most difficult
to measure and is best measured on galaxies imaged with a high S/N ratio, and
are only of limited use in this paper.  Also, the Clumpiness is highly
sensitive to resolution or seeing, although the ACS imaging we use at these
redshifts is within the range where values can be properly compared to 
nearby calibration data sets (Conselice 2003).

\subsubsection{Gini}

The Gini coefficient ($G$) is defined through the Lorentz curve of 
the light distribution and does
not depend on a predefined central position, thus distinguishing it from the
concentration parameter. If $G$ is zero, the galaxy has a uniform surface
brightness, whereas if a single pixel contains all the flux, $G$ is unity. To
calculate $G$ efficiently, the pixel flux (or stellar mass) values, 
$f_{i}$, are
sorted into increasing order and the following formula is used:

\begin{equation}
G = \frac{1}{\vert \bar{f} \vert n\left(n-1\right)} \sum_{i}^n \left(2i - n -
1\right) \vert f_{i} \vert
\label{eq:gini}
\end{equation}

\noindent where $n$ is the total number of pixels in the galaxy (Lotz \etal
2004, 2008).

\subsubsection{M$_{20}$}

M$_{20}$ is anti-correlated with concentration such that galaxies with high
concentration have low M$_{20}$ values. M$_{20}$ is a direct tracer of the
brightest regions in a galaxy, but does not require a pre-determined central
coordinate, rather this is calculated to be the location of minimal
M$_{20}$. M$_{20}$ is more sensitive to merger signatures than concentration
and is normalised by the total moment of the galaxy. It is defined as

\begin{equation}
M_{20} \equiv {\rm log} 10 \left(\frac{\sum_{i} M_{i}}{M_{tot}}\right), {\rm while} \sum_{i} f_{i} < 0.2f_{tot}
\label{Eq:M20}
\end{equation}

\begin{equation}
M_{tot} = \sum_{i}^n M_{i} = \sum_{i}^n f_{i}[(x_{i} - x_{c})^2 + (y_{i} - y_{c})^2]
\label{Eq:Mtot}
\end{equation}

\noindent where $x_{c}$, $y_{c}$ is the galaxy centre (Lotz \etal 2004).

\subsection{Simulations of CAS parameters}

One issue that we must address within this paper is how well we are measuring structure on
the stellar mass
maps for our sample of galaxies, and importantly how the measurement of stellar mass structure
would change with redshift due only to cosmological effects. This is critical if we are
to make any inferences of the evolution of galaxies in terms of their stellar mass distribution 
over time.  This issue
has been addressed before for visual images of galaxies in Conselice (2003) and in Beauvais 
\& Bothun (1999) for velocity fields in spiral galaxies. Here we address it for the stellar
mass maps.

We simulate how our stellar mass map creation process, and CAS measurements would vary as a function of
redshift due to a decrease in the signal to noise and resolution. We carry out this process
in the same way we have for the actual galaxies in the GOODS fields that we analyse in
this paper. We take a sample of 82
nearby galaxies of various types - ellipticals, spirals, and irregulars, convert their
flux values to stellar mass.  We also go through the entire process we do on GOODS, including
setting equal to zero the background values with negative stellar masses.  

This stellar mass conversion is done after the galaxy is redshifted, so that we are mimicking as
much as possible how the real observations and measurements are done.  The simulation itself
is done by simulating each image at some redshift z$_1$ to how it would appear at z$_2$, with z$_2 >$ z$_1$.
In our case z$_{1} \sim 0$.  When carrying out these simulations of placing lower redshift galaxies
to high redshifts we calculate first the rebinning factor, b, which is the reduction in apparent size of 
a galaxy's image when viewed at higher redshift.  The other major factor is  the relative amounts of 
flux from the sky and galaxy, and the noise produced from the galaxy, sky, dark current, and imaging instrument
(e.g., read noise) from ACS.   The process and details for how these simulations are done can be found
in Conselice (2003).  
 
In summary, the surface brightness of the simulated image must be reduced such that the equation,

\begin{equation}
4\pi \alpha_{z_1} N_{z_1} p_{z_1} (1+z_{1}) = 4\pi \alpha_{z_2} N_{z_2} p_{z_2} 
(1+z_{2}) \frac{\Delta \lambda_{z_1}}{\Delta \lambda_{z_2}},
\end{equation}

\noindent holds, whereby the galaxy as observed in one filter, and is simulated in
another with central rest-frame
wavelengths of $\lambda_{z_1}$ and $\lambda_{z_2}$ and widths 
$\Delta \lambda_{z_1}$ and $\Delta \lambda_{z_1}$.  In the above equation
$N_{z}$ is the total number of pixels within the galaxy at $z$ and p$_{z}$ is the average ADU counts 
per pixel.   The calibration constant
$\alpha_{z}$ is in units of erg s$^{-1}$ cm$^{-2}$ \AA$^{-1}$ ADU$^{-1}$.

A sky background, and noise from this backgrounds is added to these images by 
$B_{z} \times t_{z}$, where $B_{z}$ is background flux in units of
in units of ADU s$^{-1}$. For ACS simulations we take these values from the measured background
based on GOODS imaging (Giavalisco et al. 2004) checked to be consistent with
the values from the  ACS handbook.  Other noise effects are then added, including 
read-noise scaled for the number of read-outs, dark current and photon noise from the background.  
Our resulting images are then smoothed by the ACS PSF as generated by Tiny-Tim (Kriss et al. 2001),
although using PSFs measured from stars give the exact same results. 

From these simulated images at redshifts z = 0.5,1 and 2, we then reconstruct the stellar mass
images in the same way as we do for our original galaxies, and then measure the CAS parameters on
these stellar mass images.  The results of this are shown in Figure~\ref{fig:sim} for the concentration
and asymmetry parameters.  The M$_{20}$ and Gini parameters however have similar behaviours,
as do the galaxy half-light radii.
We find that the clumpiness indices have a similar average difference, but with a much larger
scatter.  
  
What we find overall is that the trend with redshift is such that the concentration and asymmetry 
parameters on average do not change significantly when measured in stellar mass (Figure ~\ref{fig:sim}).  
We show in the right panel of Figure ~\ref{fig:sim} the change in C and A when divided into
early/late/peculiar types.  Again, on average there is not a significant change with redshift,
although individual galaxies clearly can have significant differences at higher redshift.  

We also investigate how our parameters change at each redshift between the stellar mass
and the visual images after the simulation.  We find very little difference except that at the highest redshifts
of our simulations, at $z = 2$, we find that the average change in the asymmetry parameter
changes $\delta A = (A_{\rm opt} - A_{\rm mass}) \sim -0.1$, such that the stellar mass image
is more asymmetric than the optical light one.   
We also investigate the same quantities that we use in later figures,
finding that twice the difference in the asymmetries of the optical and stellar mass
image, divided by the sum of these, increases steadily until it reaches a values of
$-1.1$ at $z = 2$ similar to what we see later in this paper for the actual values.  However
at $z = 1$ we find that this change is less and closer to $-0.1$.

\section{Analysis}\label{sec:results}

The following sections describes the analysis of our sample, 
and what we can learn from examining their
structures in stellar mass maps, and how this evolves
over time.  We first examine the
distribution of visual classifications for our sample and
their global colours.  We later describe the stellar mass maps we
construct for these galaxies, and then finally present a 
structural analysis of these systems based on the distribution
of their stellar mass.

\subsection{Global Properties of the Sample}

In Figure ~\ref{fig:CMD} we show the observed colour magnitude diagram for our
sample.  Each morphological class is plotted as a different symbol; solid red  
circles represent the cE's, open red circles show the pE galaxies and the  
E galaxies are represented by red stars. Solid blue squares show the eS  
and open blue squares the lS systems. Edge-on disk galaxies are represented by black 
diagonal crosses and solid green triangles show the Pec systems. 
pM and M galaxies are represented by open green triangles and open 
green stars respectively. This key is used throughout the rest of the 
paper in the various figures.

Based on this, we find that our classified early-type galaxies are on average 
redder and brighter than the peculiars and pM/M systems, while the spirals 
tend to be bluer across a range of brightnesses, but the differences are not 
so clear as would be expected for a sample at low-$z$ only.
 This colour-magnitude diagram does not show such an obvious morphological 
sequence as is seen for purely Hubble type galaxies in the nearby universe
(e.g., Conselice 2006b). 

The relation between $(V_{606} - z_{850})$ colour and redshift is shown in
Figure ~\ref{fig:colz}. The symbols are the same as those in Fig.
~\ref{fig:CMD}. As can be seen, there is a general trend for all types to 
become redder with increasing redshift, which is an expected feature of redshift. 
The early types are reddest across the whole redshift range, followed by the 
eS and lS galaxies, with the Pec/pM/M systems the bluest across all 
redshifts. The compact ellipticals are blue and 
faint. Note however that some pM and eS galaxies are quite red.  
What this reveals is that there is a decoupled relation between the 
colour and morphologies of galaxies not seen in nearby galaxies where
this correlation is strong. A specific morphological type as measured
visually at high-z cannot be used to predict the colour of the galaxy.

\subsection{Stellar Mass and Mass to Light Ratio Maps}

By calculating stellar mass to light ratios and the stellar mass within
each pixel of each galaxy image in our sample, we reconstruct the image of
each galaxy in terms of stellar mass and mass-to-light
ratio. We do this for each galaxy, using the method described in LCM07, 
and present five representative examples for each classification in Figures 
~\ref{fig:massmap_cE} to ~\ref{fig:massmap_M}. This figures present galaxy
images in $z_{850}$ (left), stellar mass (centre) and $(M/L)_B$ (right), which
we discuss below. The images are grey-scaled such that the darkest pixels
represent those brightest in $z_{850}$, and those that are most
massive. In the mass-to-light image, the whitest pixels are those with the
highest ($M/L$), or redder in colour.  We explain below how these maps
appear for our various types.

\subsubsection{Compact Ellipticals}

Figure ~\ref{fig:massmap_cE} shows five examples of the compact elliptical
(cE) population in the sample. These galaxies appear similar in their 
stellar mass maps as they do in their $z_{850}$ images at first
glance.  The  
mass-to-light maps are not as uniform, and there are important quantitative
differences between the stellar mass and light images. 
For example, the galaxy 18246 appears to have a relatively 
higher ($M/L$) in its core compared to its outer parts, whilst galaxy
25035 appears to have a low ($M/L$) in its centre, suggesting that it has a 
bluer core, 
likely due to star formation.  Galaxy 36503 contains a high ($M/L$) 
ring surrounding a blue core.

\subsubsection{Peculiar Ellipticals}

Five examples of the peculiar elliptical (pE) galaxies are displayed in
Fig. ~\ref{fig:massmap_pE}. These galaxies appear  more diffuse in stellar 
mass than in $z_{850}$, in part due to the loss in contrast in the stellar 
mass maps, but also due to their blue inner colours 
(Fig. ~\ref{fig:massmap_pE}). As can be seen in the ($M/L$) ratio maps of 
these galaxies, they have blue cores, and the effect is most pronounced in 
22932 and 27429.  These galaxies also have blue overall 
colours compared to pure E galaxies (Fig. ~\ref{fig:colz}).   These peculiar 
ellipticals have been seen and studied before in papers such as Conselice 
et al. (2007).  They generally have a high asymmetry, and are found amongst 
the most massive galaxies in the universe at $z \sim 1$.  As can be seen in 
the M/L maps for these systems (Fig. ~\ref{fig:massmap_pE}), these galaxies 
also have a diversity in how young and old stellar populations, including
dust, are distributed in these systems compared to normal elliptical galaxies.

\subsubsection{Ellipticals}

Figure ~\ref{fig:massmap_E} presents five examples of early-types from our
sample. These galaxies appear more diffuse and more distributed spatially
in their stellar 
mass than in the $z_{850}$ band, especially at large radii. These galaxies 
have larger ($M/L$) ratios in their centres, except for 36419, which is blue, 
and has a more complicated structure in ($M/L$) than the others. Overall, 
the E galaxies morphologically  have very similar structures in stellar mass 
and $z_{850}$.

\subsubsection{Spirals}

Figures ~\ref{fig:massmap_eS} and ~\ref{fig:massmap_lS} show the $z_{850}$,
stellar mass and ($M/L$) maps for the early-type, and late-type spirals. 
As found in LCM07,  structures within discs, including prominent spiral 
arms, are often (but not always) smoothed out in stellar mass. This 
is true for early types as  well as late-type spirals, as can be seen for
example in
galaxy 25465 (Fig. ~\ref{fig:massmap_eS}). Inspection of the ($M/L$) maps 
reveal that the discs of many spirals are bluer than their bulges, as expected.

The mass-to-light ratio maps of the edge on galaxies 
(Fig. ~\ref{fig:massmap_EO}) are mostly homogeneous in structure. However, 
the maps of 19280, 39312 and 49722 all show some patches of low $M/L$. 
Indeed, 19280 appears to be blue across the whole image, which would 
imply that the dust content of this galaxy is low, which is unusual for 
edge-on galaxies.

\subsubsection{Peculiars \& Mergers}

Figure ~\ref{fig:massmap_Pec} shows the F814W images, stellar mass maps and M/L 
ratio maps for the Peculiar galaxies within our sample. The Peculiar 
galaxies 22690 and 25584 (Fig. ~\ref{fig:massmap_Pec}) appear to contain
 objects near the primary galaxy, but in both cases it is the larger galaxies on the
right that is the target.  Although 22690 looks disturbed in $z_{850}$, 
the effect of smoothing in the stellar mass map is also seen here. 
This indicates that disturbed regions of the galaxy are due to star 
forming regions. The $z_{850}$ image of
22690 does not have the regular morphology of a nearby spiral galaxy, although
the stellar mass map has a structure one would expect of such a 
galaxy, having a central bulge surrounded by a smooth disc. The ($M/L$) 
map also shows a red central region surrounded by bluer pixels.

The mass-to-light maps of the Peculiars generally show that they are blue and
also difficult to see in stellar mass, due to their low mass to light ratio. 
This suggests that star formation
itself might be difficult to trace in stellar mass. However, the stellar mass 
in the pM galaxy
18917 (Fig. ~\ref{fig:massmap_pM}) traces the light in the galaxy, despite
having a low ($M/L$).  Note also that the peculiars are often peculiar spirals,
in the sense that they look like nearly normal spirals, but with some
peculiar features. Often these peculiars are quite small as well, and many of 
them are very likely spirals in some type of formation.


The merging galaxy system 21839 (Fig. ~\ref{fig:massmap_M}) is similar in 
stellar mass to its $z_{850}$ image, but possesses a more intricate structure 
in ($M/L$). The mass-to-light map shows the smaller merging object to be blue,
whereas the brighter galaxy appears more red but with blue regions in its core.
 
The system 29800 (Fig. ~\ref{fig:massmap_M}), whilst clearly appearing to
be a
merger between two galaxies of approximately equal brightness in $z_{850}$,
would likely be classified as a disc galaxy in stellar mass. The bottom of 
the two bright nuclei has disappeared completely in the stellar mass image, 
although it is visible as a low ($M/L$) patch in the mass-to-light map.

\subsection{Comparison Between Galaxy Properties in Mass and Light}

In this section we compare how the distribution of stellar mass in a galaxy
compares to the distribution of light as seen in the $z_{850}$ ACS imaging.
One of our goals is to determine how appropriate studies in $z_{850}$ and 
similar red bands are for measuring the mass content and structure of a galaxy, 
and how stellar mass quantitative morphologies differ from those measured in light. 

In this section we investigate the relationship between galaxy properties 
in stellar mass and $z_{850}$-band light. For galaxy size (as  
measured by the half-light radius) and the $CAS$ parameters we  
plot the normalised difference between $z_{850}$ and stellar 
mass versus the mean value of $z_{850}$ and stellar mass as a representative  
figure to determine how these various quantities change.

\subsubsection{Galaxy Half-light and Half-mass Radii}

One of the basic features of a galaxy is its size, which in this paper we
quantitatively mean the half-light radii.  Galaxy half-light radii are 
found to strongly evolve with time, such that galaxies at higher redshifts have
a more compact structure (e.g., Ferguson et al. 2004; Trujillo et al. 2007; 
Buitrago et al. 2008; Carrasco et al. 2010).
However, every size measurement has been carried out through measurements
of light, and it is desirable to determine how the half-mass radii of galaxies measured
in stellar mass maps compares with half-light radii measured using light.

As such, we have calculated half-light (or half-mass) radii ($Re$), in kpc, 
for our sample in both stellar mass and $z_{850}$ light, and investigate whether 
half-light/mass radii are comparable. 
We define the normalised size difference as 

$$2[R_{\rm e}(z_{850})-R_{\rm e}(M_{*})]/[R_{\rm e}(z_{850})+R_{\rm e}(M_{*})],$$  

\noindent and the average size as 

$$\frac{1}{2}[R_{\rm e}(z_{850})+R_{\rm e}(M_{*})].$$ 

\begin{figure}
\centering
\includegraphics[angle=0, width=90mm, height=90mm]{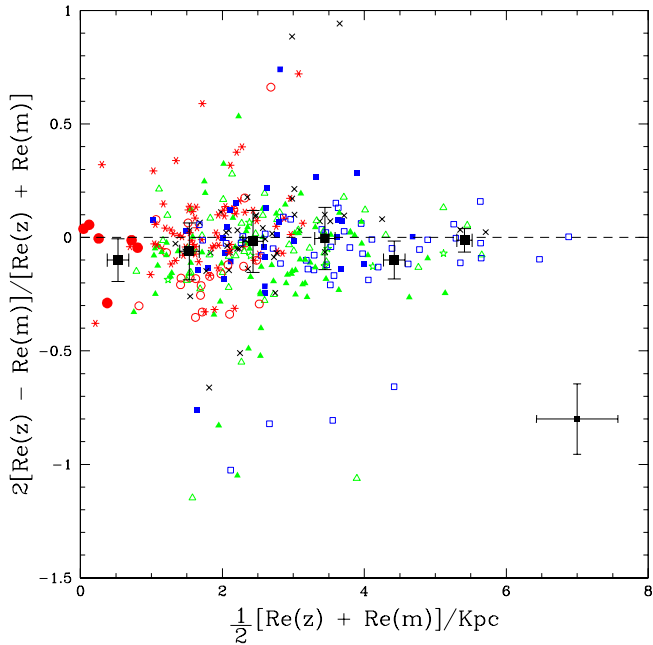}
\caption{Normalised effective-radius difference vs. mean 
effective radius of the sample in $z_{850}$
  and stellar mass. The error bar to the right of the plot shows the average
  dispersion for the whole sample. The large black squares denote the average
  values in equally spaced bins of mean size, with error bars showing the
dispersions of these values.  This convention
  is used in all of the parameter comparison plots.  The symbols plotted
here are the same as in Figure~12.}
\label{fig:size_cf}
\end{figure}

\noindent We use these two calculations so as to avoid
biasing the analysis when the values become very
small in either the stellar mass image, or in the $z_{850}$ band.
We plot the relation between these values in  
Figure ~\ref{fig:size_cf}. The horizontal dashed line marks the position
of equal size ($Re(z_{850}) = Re(M_{*})$). The symbols are the same as in 
Figure~\ref{fig:CMD}, and the average dispersion for the sample is
represented by the black square points on the right of the plot. The larger
black square points in the figure show the average values and the
measured dispersion for
the ranges of $\frac{1}{2}\left( Re(z_{850}) + Re(M_{*}) \right)$ between 
zero and one kpc, one and two kpc, up to six kpc. This convention is 
used for each of the parameter comparisons in 
Sections ~\ref{sec:conc_cf} to ~\ref{sec:M20_cf}.

We note that there is a steady progression for increasing average galaxy size
with morphology,  from the compact ellipticals, early types
and early spirals to the late-type spirals, with the peculiars and edge-on
galaxies being more randomly distributed. The points are scattered about the
line of equality such that nearly half (42 percent) of the galaxies have 
$Re(z_{850}) > Re(M_{*})$. There is, therefore, a slight tendency for 
sizes in masses to be higher, but this does not vary significantly with 
average galaxy size and type (see Table~2).

There is no clear tendency for any particular morphological type to be larger
in either the $z_{850}$ or stellar mass image, which suggests that there 
is no bias introduced 
by using measurements in $z_{850}$ band data for size measurements 
(Trujillo \etal  2007; Buitrago et al. 2008).  Table ~\ref{tab:size_cf}.  
shows that the average difference between the sizes in the 
stellar mass maps and the $z_{850}$-band image is essentially zero, 
demonstrating that the measurements of half-radii in light does not differ 
significantly from the measurements in the distribution of stellar mass.

We also examine the normalised difference in size
against ($V_{606}$ - $z_{850}$) colour to test whether there is a tendency for
bluer galaxies to have larger radii in stellar mass than in light. This
is what we might expect to find if star formation dominated the light near the
centre of the galaxy. While, on average, colour does not change with size
difference, we find that extreme half-radii differences between stellar
mass and $z_{850}$ are mostly in blue systems.

\begin{table}
\centering
\begin{tabular}{c  r  r}
\hline \hline
\\
Type & Mean Size & Normalised Size Difference \\
       &   (kpc)                               &         \\                
\\
\hline
cE   &   $0.6 \pm(0.3)$  &  $0.14 \pm(0.24)$   \\
pE   &   $1.8 \pm(0.2)$   &  $-0.11 \pm(0.11)$     \\
E    &   $1.8 \pm(0.3)$   &  $0.05 \pm(0.11)$   \\
eS   &   $2.6 \pm(0.4)$   &  $0.00 \pm(0.12)$     \\
lS   &   $3.7 \pm(0.6)$   &  $-0.10 \pm(0.12)$     \\
EO   &   $2.7 \pm(0.5)$  &   $0.02 \pm(0.13)$    \\
Pec  &   $2.6 \pm(0.5)$   &  $-0.09 \pm(0.13)$     \\
pM   &   $2.7 \pm(0.6)$   &  $-0.07 \pm(0.13)$     \\
M    &   $3.0 \pm(0.6)$  &   $-0.08 \pm(0.03)$     \\ 
\hline
\end{tabular}
\caption{Comparisons between mean $z_{850}$-stellar mass half-mass
radius ($R_{\rm e}$) and normalised size difference in $z_{850}$ and stellar 
mass, organised by morphological type.} 
\label{tab:size_cf}
\end{table}

\subsubsection{Concentration}\label{sec:conc_cf}

\begin{figure}
\centering
\includegraphics[angle=0, width=90mm, height=90mm]{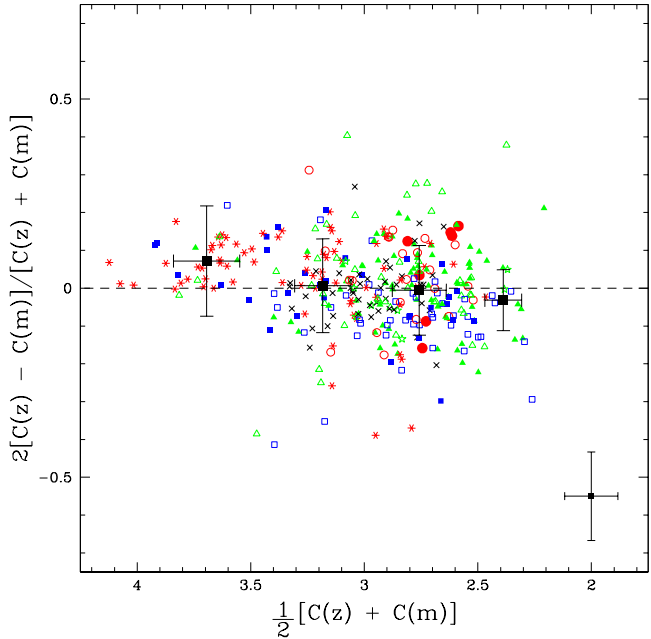}
\caption{Comparison between the concentration parameter in the $z_{850}$
band  and within
 stellar mass maps. The error bar to the right of the plot shows the average
  dispersion for the whole sample. The large black squares denote the average
  values in equally spaced bins of mean concentration, plus dispersions.  The
symbols used here are the same as in Figure~12.}
\label{fig:conc_cf}
\end{figure}

Analogous to the galaxy size comparison in the previous section, we compare
the concentration parameter in $z_{850}$ and stellar mass images, as plotted 
in Figure~\ref{fig:conc_cf}.  The
dashed line shows $C(z_{850}) = C(M_{*})$ and the black square points
represent the average values in equally spaced bins of average concentration.

The average concentration, in both $z_{850}$ and stellar mass, is generally 
high for the early types, and lower in the late-types and peculiars, 
with the early-type spirals being spread between low and high values. 
The compact elliptical galaxies have a low average concentration, due 
to the light being spread evenly across a galaxy's pixels.

The average trend shows that the ratio $\frac{R_{80}}{R_{20}}$ is slightly
lower in the stellar mass maps for the early types compared to the $z_{850}$
band concentration. We find that in general R$_{80}$ is lower
and R$_{20}$ is higher, compared to those in the $z_{850}$-band
image, although the effect is mostly due to a smaller R$_{80}$ although
this effects is quite small.  It is possible that 
stellar mass is less centrally concentrated towards the centre of 
the galaxy in the early-types (raising
R$_{20}$), and/or the stellar mass is more diffuse at larger radii than
the distribution of light (lowering
R$_{80}$). It is also the case that for many galaxies
more star formation is seen in the outer regions of the galaxies 
in the $z_{850}$-band, but these pixels are less dominant in stellar mass, 
raising the value of R$_{80}$ creating higher values of $C$ as seen.

\subsubsection{Asymmetry}\label{sec:asym_cf}

\begin{figure}
\centering
\includegraphics[angle=0, width=90mm, height=90mm]{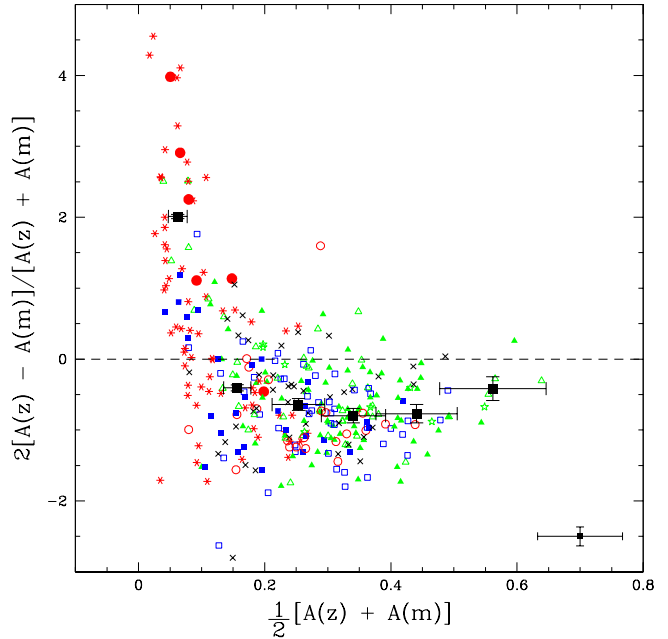}
\caption{Comparison between the asymmetry parameter in $z_{850}$ and stellar mass. The 
points and conventions are the same as in Figure~12.  Note that the asymmetry
in the stellar mass is lower than the $z-$band for only the early types. This
is due to the method of measuring the asymmetry, where the sky background
is handled in a different way than for imaging, resulting in higher values
for later type galaxies (\S 4.3.3). }
\label{fig:asym_cf}
\end{figure}

Before we discuss the comparison of our stellar mass map asymmetry values
to the $z_{850}$-band asymmetries, we note a few things.  
First, the lowest asymmetry values for the sample are negative 
in all bands, mainly due to the background correction which for very
symmetric galaxies will sometimes be larger than the asymmetry in the galaxy
itself.   This has the effect, 
as seen in Figure ~\ref{fig:asym_cf}, of skewing the low average asymmetry 
values differences for the smallest stellar mass measured asymmetries. 
As can be seen in Figure~16, except for these galaxies with low asymmetry 
values, there is a clear tendency for galaxies to be more asymmetric in 
stellar mass than in $z_{850}$, this being the case for a large
fraction of the non-early type sample. The stellar mass maps also have a 
greater spread in asymmetry values than $z_{850}$, and this can be more 
clearly seen in \S ~\ref{sec:AC}.

 However, we also note that the blue regions of galaxies are 
not so well traced by the stellar mass
maps, and often led to difficulties with the image contrast, making the galaxy
features harder to see, as in the case of the late-spiral 19535
(Fig. ~\ref{fig:massmap_lS}).  Further galaxies, such as 38722 (Figure~6), 
are clearly more asymmetric and lopsided when viewed in the stellar mass 
band.  In this case, it appears that one spiral arm remains while the other 
disappears.

There are several reasons why the asymmetry value in the 
stellar mass maps is higher than in the images.  First, 
when calculating $A$, noise tends to be magnified due to the process of
subtracting images.   The simulations discussed in \S 3.4 show that
there is a tendency for the average galaxies to become more asymmetric
in stellar mass measurements than in light.  This can explain part of
this, but we find that the average difference of around $-1$ in the
relative difference is too high to be accounted for solely by these
redshift effects.  We discuss some of the reasons for why the
asymmetries will be higher in the stellar mass maps than in light.

When calculating stellar masses for each pixel we assume
that the $M/L$ ratio is approximately constant in the surrounding 
pixels. This should be the case for early type galaxies,
especially, due to their uniform stellar populations. To test how much this
variance in $M/L$ could be affecting the $A$ values, we have measured
the $M/L$ pixel variance in a typical early type galaxy (30976,
Fig. ~\ref{fig:massmap_E}). The mean $M/L$ for a pixel in this galaxy is
$(M/L)_{B} = 1.29$ with a typical standard deviation in the surrounding pixels
of $\sigma_{M/L} = 0.27$. Although a correction has been applied to minimise
this effect (see \S 3.3.1), it is not enough to account for
the high asymmetry signal in the spiral galaxy stellar mass maps, and thus 
this asymmetry is likely a real effect due to the nature of the calculation
of this parameter, and especially the background, as described in \S 3.3.1.

\begin{figure*}
\centering
\includegraphics[angle=0, width=188mm, height=95mm]{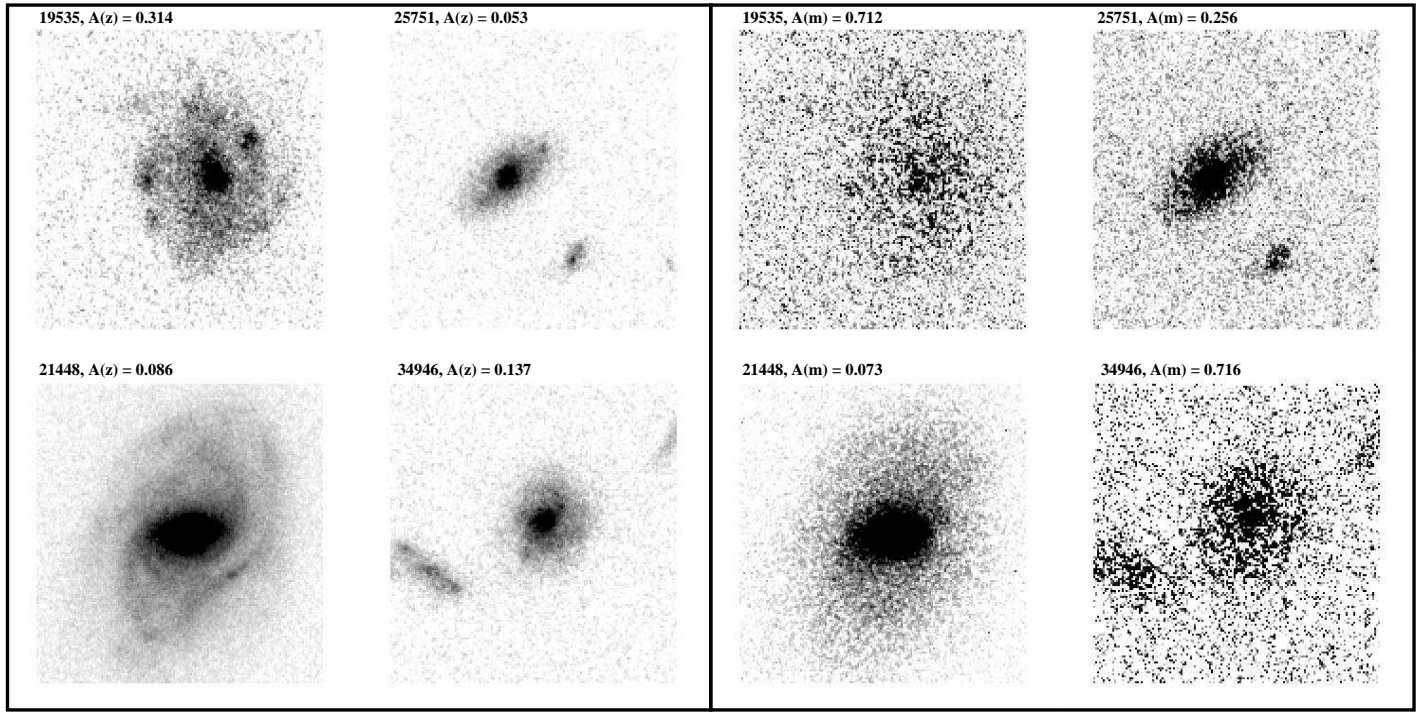}
\caption{Figure showing the $z_{850}$-band imaging (left) and
stellar mass maps (right) for four late-type spiral 
galaxies. Each galaxy is 
labelled with its ID number and $A(z_{850})$ value. Each image is 
4.5 arcseconds on each side. 
Shown at the top of each image is the ID and the asymmetry parameter
measured on the stellar mass maps to be compared with the same
value computed in the $z_{850}$ band.}
\label{fig:spirals_z}
\end{figure*}

To further understand why late-type spiral galaxies are more asymmetric 
in stellar mass than $z_{850}$, we have examined these galaxies in detail 
and present four examples of the late-type sample in both 
$z_{850}$ (Fig. ~\ref{fig:spirals_z}), and stellar mass maps.  
Two of these galaxies have low asymmetries
in both stellar mass and $z_{850}$ (25751 and 21448), and galaxy 21448 in 
particular, demonstrates the smoothing of spiral arms, which we observed 
previously in nearby galaxies (LCM08). However, two of these galaxies have 
higher asymmetries in stellar mass than in $z_{850}$ (19535 and 34946). 
Object 19535, especially, shows little relation in stellar mass to the 
$z_{850}$ image. The star forming knots and central bulge, which can be 
seen clearly in $z_{850}$, translate to large scale asymmetries in stellar 
mass. There are several possible explanations for this effect. The star 
forming regions in 19535 could truly be more massive per pixel than the 
surrounding disk, due to the high masses and densities of gas and dust 
required to form such massive star forming regions. This would lead to 
such regions not being smoothed out in stellar mass, causing higher $A(M_{*})$
values. It is also possible that what we have interpreted as star forming
knots in the disk in some cases could be minor merger signatures, that is, 
low-mass galaxies falling into the object itself.

There is also a trend with morphological-type with increasing asymmetry 
such that early types have low asymmetry and late-types higher asymmetry. 
Our findings confirm previous studies who found similar trends 
(e.g. Conselice \etal\,  2000). We have also investigated  the relation 
between asymmetry in stellar mass and $(V_{606}-z_{850})$
colour, and examine this after splitting into two
redshift bands: $z < 0.6$ and $0.6 \leq z < 1$. There is a clear trend for
galaxies with  a higher degree of asymmetry to be blue, with
$\langle(V_{606}-z_{850})\vert_{A(M_{*})<0.35}\rangle = 1.01(\pm0.20)$ and
$\langle(V_{606}-z_{850})\vert_{A(M_{*})\geq0.35}\rangle = 1.22(\pm0.27)$.  
This trend has also been 
found for  asymmetries measured in light (Conselice \etal 2003). 


\begin{figure}
\centering
\includegraphics[angle=0, width=90mm, height=90mm]{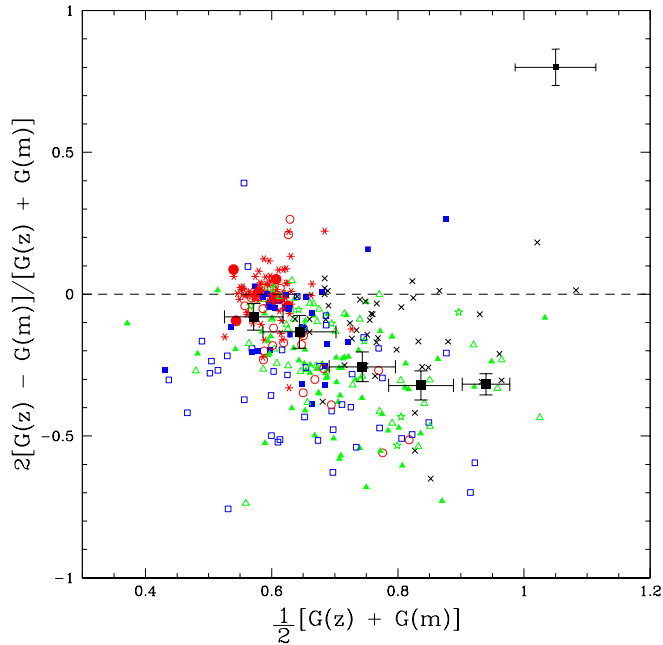}
\caption{Comparison between the Gini parameter in $z_{850}$ and stellar mass. 
The points and conventions are the same as in Figure~12.}
\label{fig:gini_cf}
\end{figure}

\begin{figure}
\centering
\includegraphics[angle=0, width=90mm, height=90mm]{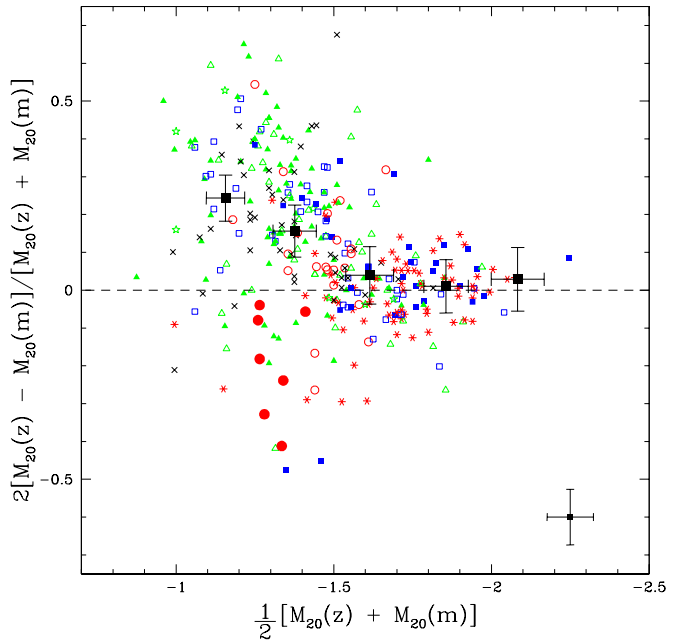}
\caption{Comparison between the M$_{20}$ parameter in $z_{850}$ and stellar mass. The points and conventions are the same as in Figure~12.   Because of the inverse nature of M$_{20}$, galaxies with positive differences on
the y-axis are those with more `concentrated' light profiles than in the
stellar mass maps.}
\label{fig:m20_cf}
\end{figure}

\subsubsection{Clumpiness}

The clumpiness values in stellar mass maps have a higher degree of 
uncertainty, as discussed in Section 3.3.3.
Overall, we find a
trend for clumpiness to be greater in stellar mass than $z_{850}$, and this is the 
case for 91 percent of the sample. However for larger values of average $S$, 
there is a trend such that $S(M_{*}) \rightarrow S(z_{850})$. 

\begin{figure*}
\centering
\includegraphics[angle=0, width=148mm, height=148mm]{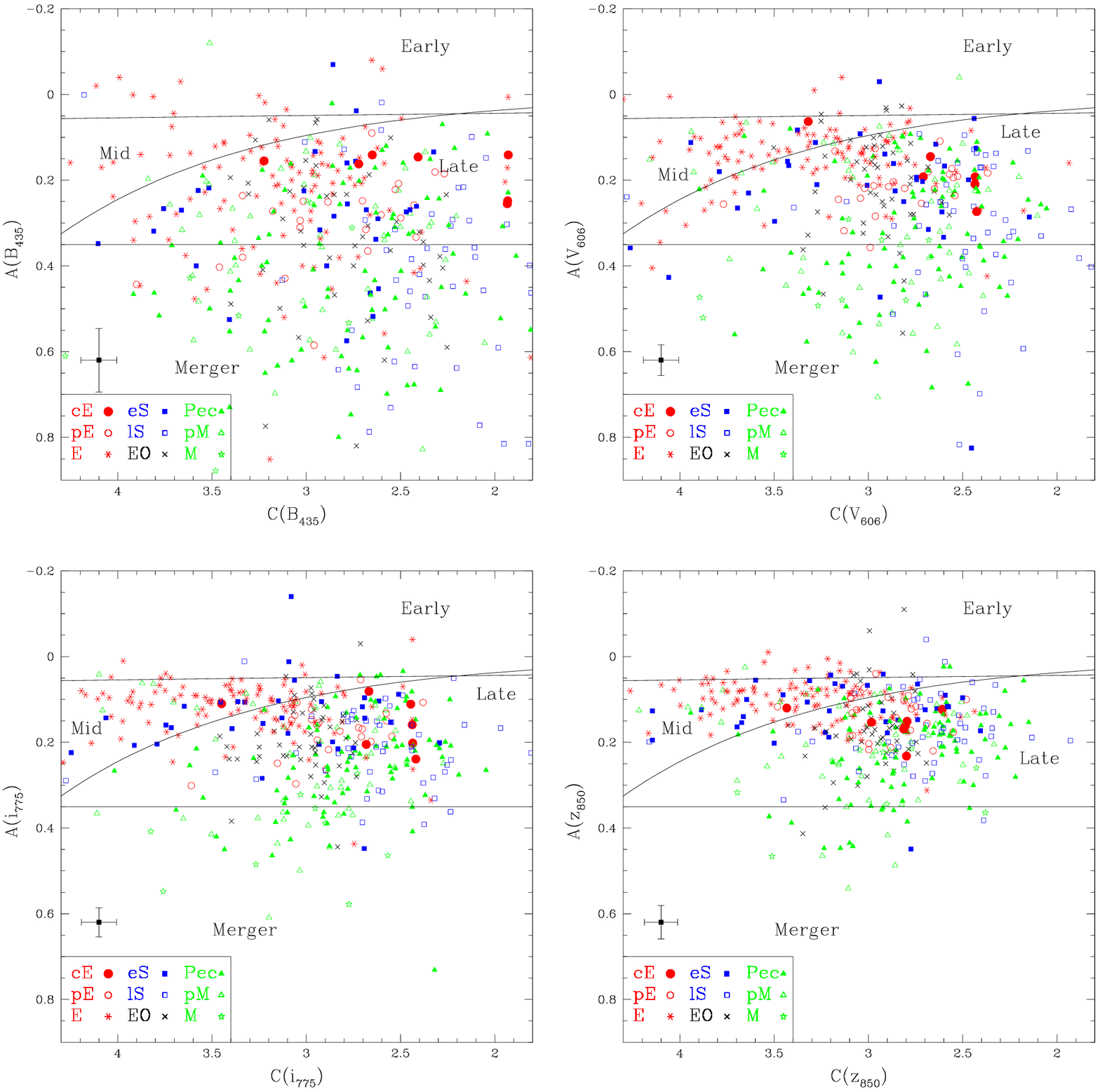}
\caption{Asymmetry vs. concentration in the $B_{435}$, $V_{606}$, $i_{775}$
  and $z_{850}$ bands. Solid lines show the classification
criteria of Conselice (2003). 
The point
  in the bottom left segment of the plot shows the average error bar for the
  whole sample. }
\label{fig:AC_bviz}
\end{figure*}

\begin{figure}
\centering
\includegraphics[angle=0, width=90mm, height=90mm]{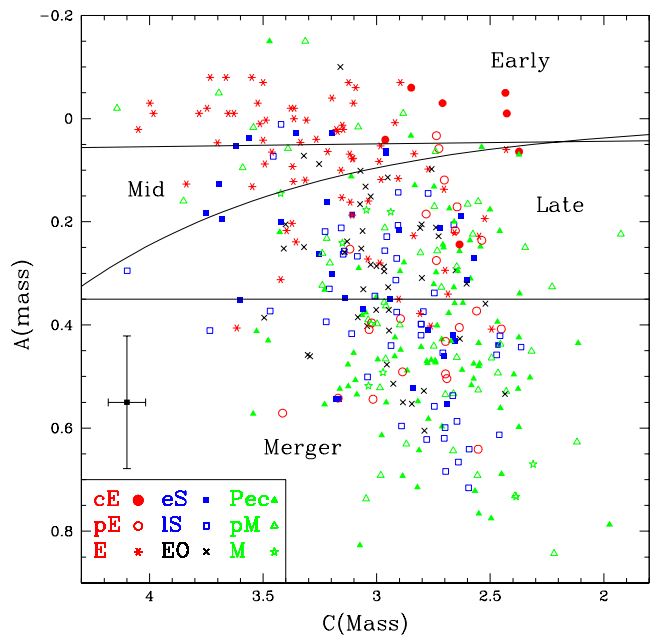}
\caption{The asymmetry-concentration plane for the stellar mass maps.  The 
plotting  convention is the same as for Figure~12.}
\label{fig:AC_mass}
\end{figure}

The clumpiness parameter is similar to asymmetry, but picks out small 
scale features, such as compact star clusters, as opposed to large scale 
asymmetries. It is, therefore, not surprising that uncertainties in $M/L$ 
cause a large number of the sample to have $S(M_{*}) > S(_{850})$, as 
variations in $M/L$ from pixel to pixel filter through in the 
calculation of $S(M_{*})$. Such variations affect $A(M_{*})$ and $S(M_{*})$ 
more than 
the other parameters, as these calculations involve the subtraction of 
images, which amplifies the $M/L$ variations and large
variations of level within the background.


\subsubsection{Gini Index}

Figure ~\ref{fig:gini_cf} shows the comparison between the values of the Gini
parameter in stellar mass and the $z_{850}$ band. 
Although the early-types are more strongly clustered around equality, 
$G(M_{*}) = G(z_{850})$, there is a clear trend for the sample to have 
higher $G$ in stellar mass, with $G(M_{*}) > G(z_{850})$ in non-early types
within our sample. This comparison of $G(z_{850})$ and $G(M_{*})$ 
displays a greater separation between types. The late-type spirals 
show an approximately even 
spread in Gini values, from ~0.45 to ~0.93, but all of these have a higher 
difference between $z_{850}$ and stellar mass than the early-types
and compact ellipticals. 

Figure ~\ref{fig:gini_cf} reveals that the Gini index is higher within the stellar 
mass maps than in 
the $z_{850}$ band, except for early-type galaxies, indicating 
that most of the 
stellar mass in later morphological types is contained within fewer pixels
within the stellar mass image. 
This is a significant difference between the early and late-types in our 
sample.  The reason for this is that the bright blue
regions of these galaxies vanish as the M/L ratio is inversely proportional
to L.  This creates effectively a mass distribution with
a larger fraction of the mass contained within fewer pixels.
For example the bulges become more prominent for the late-type disks while the
arms vanish.  The bulges in these systems become more prominent and this
will rise the value of the Gini index.

What can often been seen in stellar mass maps  of late-types and 
mergers/peculiars is that these bright outer regions vanish, 
leaving only the central part of the galaxy  including most objects in 
Figure~6 which loose their outer parts
and appear almost as early-types.  Figure~18 also shows that the 
Gini index for the early-types in 
stellar mass and light are similar, likely because of the 
small variation in the M/L ratios of the various pixels. 

\subsubsection{M$_{20}$}\label{sec:M20_cf}

We show the differences between the M$_{20}$ index in the stellar mass
maps and the $z_{850}$ band in Figure ~\ref{fig:m20_cf}.
Due to the reversed parity in the M$_{20}$ values, points which lie above the
dashed line in Figure ~\ref{fig:m20_cf} have M$_{20}(M_{*}) >$
M$_{20}(z_{850})$, that is, the stellar mass is more more diffuse than the
light for most systems.   This would change however if we used
a smaller radius than the standard definition.  Overall, M$_{20}(M_{*})$ is higher
(less negative) than M$_{20}(z_{850})$ for most of the sample.

The M$_{20}$ parameter traces the brightest 20 percent of the flux in the 
galaxy, or the greatest 20 percent of the stellar mass. This parameter is  
heavily weighted by the spatial distribution of the most luminous, or most 
massive, pixels, but is normalised to remove any dependence on galaxy size 
and total flux/stellar mass. M$_{20}$, like $G$, is also not dependent on a 
fixed, pre-determined centre.

Therefore, if stellar mass exactly follows light within a galaxy we would 
expect that
M$_{20}(M_{*})$ = M$_{20}(z_{850})$. Figure ~\ref{fig:m20_cf} shows that, while
this is approximately the case for early-type galaxies and early spirals, the
same does not apply for peculiars, late-type spirals and edge-on discs. The
compact ellipticals especially differ from this trend, 
having M$_{20}(M_{*})$ $<$ M$_{20}$($z_{850}$).  This is certainly due to 
the fact that these compact ellipticals have blue cores which render their 
stellar mass maps more compact in the centre than for galaxies with 
redder cores.

\subsection{Stellar Mass Structure}

\subsubsection{Concentration vs. Asymmetry} \label{sec:AC}

We examine how galaxies fall in the classic 
concentration-asymmetry plane (e.g., Conselice et al. 2000) to determine 
if different galaxy types, as determine visually can be better separated 
at different wavelengths in this parameter space.

We plot the relation between asymmetry and concentration in light for the
sample in Fig.~\ref{fig:AC_bviz} and in stellar mass in 
Fig. ~\ref{fig:AC_mass}.  Over-plotted are the classification  criteria 
of Conselice (2003) and Bershady et al. (2000). Galaxies with 
$A>0.35$ are classified as mergers, 
while  those which lie above the top line are early-types. 
Galaxies to  the left of the middle line in Figure~20 \& 21, 
are labelled 
mid-types and those to right of this line are classified as  late-types. 




These criteria for identifying mergers and 
late-types appear  more appropriate for our sample in the $i_{775}$ and 
$z_{850}$ bands, while galaxies  are more mixed in these regions in stellar 
mass, having many more
galaxies with high asymmetries, and spirals are not easily distinguished from
merging systems.  However, the uniqueness of the early-type region is more
obvious in stellar mass
maps, with most E  galaxies in the $i_{775}$ and $z_{850}$ band plots 
falling into the mid-type region (Fig. ~\ref{fig:AC_mass}).

\begin{figure*}
\centering
\includegraphics[angle=0, width=148mm, height=148mm]{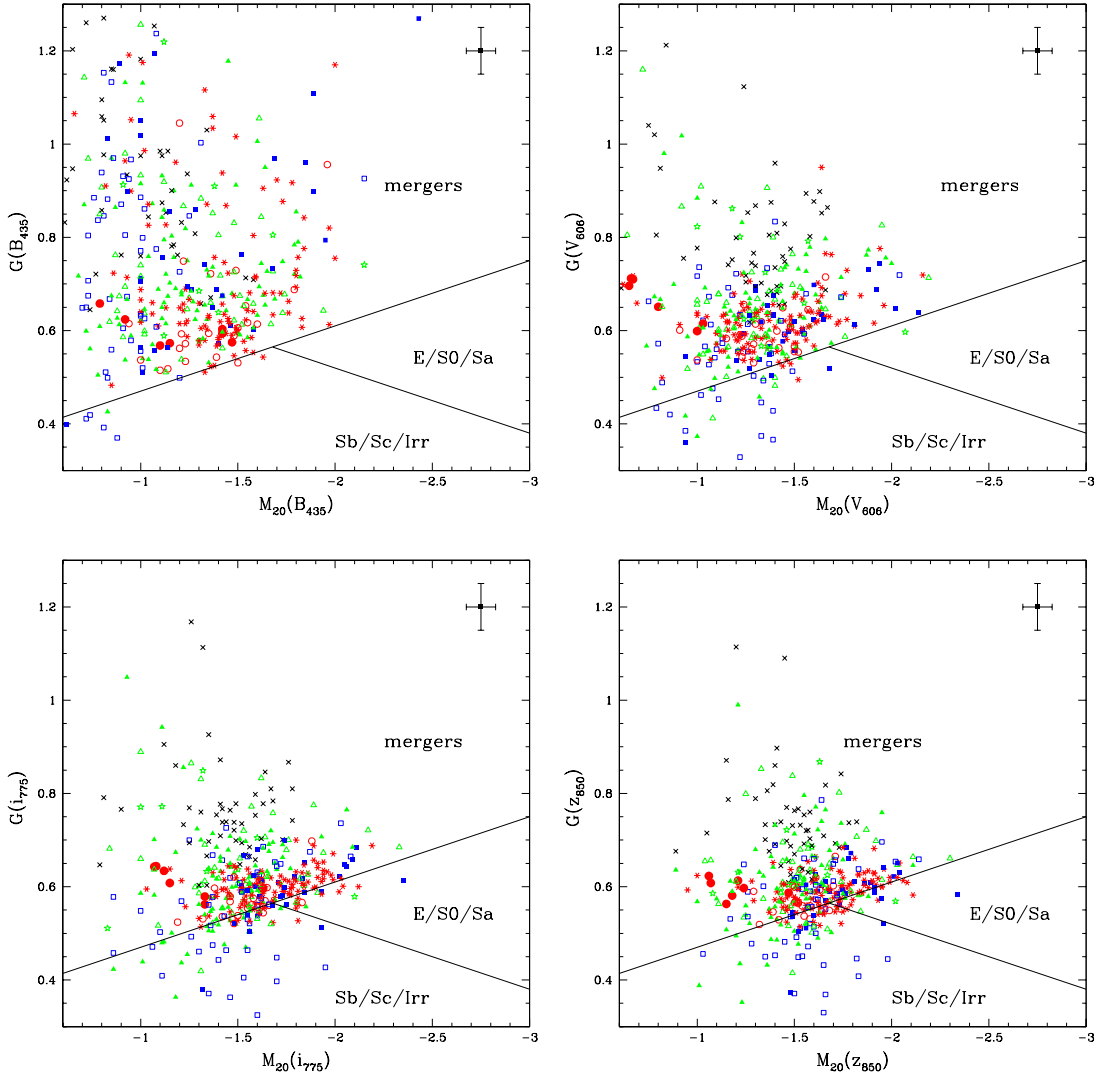}
\caption{Gini-M$_{20}$ for the $B_{435}$, $V_{606}$, $i_{775}$ and $z_{850}$
bands. The solid lines mark the classification criteria of Lotz \etal  (2004;
Equations ~\ref{eq:lotz_mergerline} and  ~\ref{eq:lotz_normalline}).}
\label{fig:gm20_bviz}
\end{figure*}

We find that for our sample, $z_{850}$ is the best band in which to measure
these parameters for separating morphological types from each other. 
It can also be seen in Fig. ~\ref{fig:AC_bviz}  that the $A$-$C$ 
relation breaks down when viewed in bluer bands, as also shown for nearby 
galaxies in Taylor-Mager et al. (2007).  We find that the scatter in $A$-$C$ increases for all
galaxy types towards bluer wavelengths, the effect is more pronounced for 
late-type morphologies.

\begin{figure}
\centering
\includegraphics[angle=0, width=90mm, height=160mm]{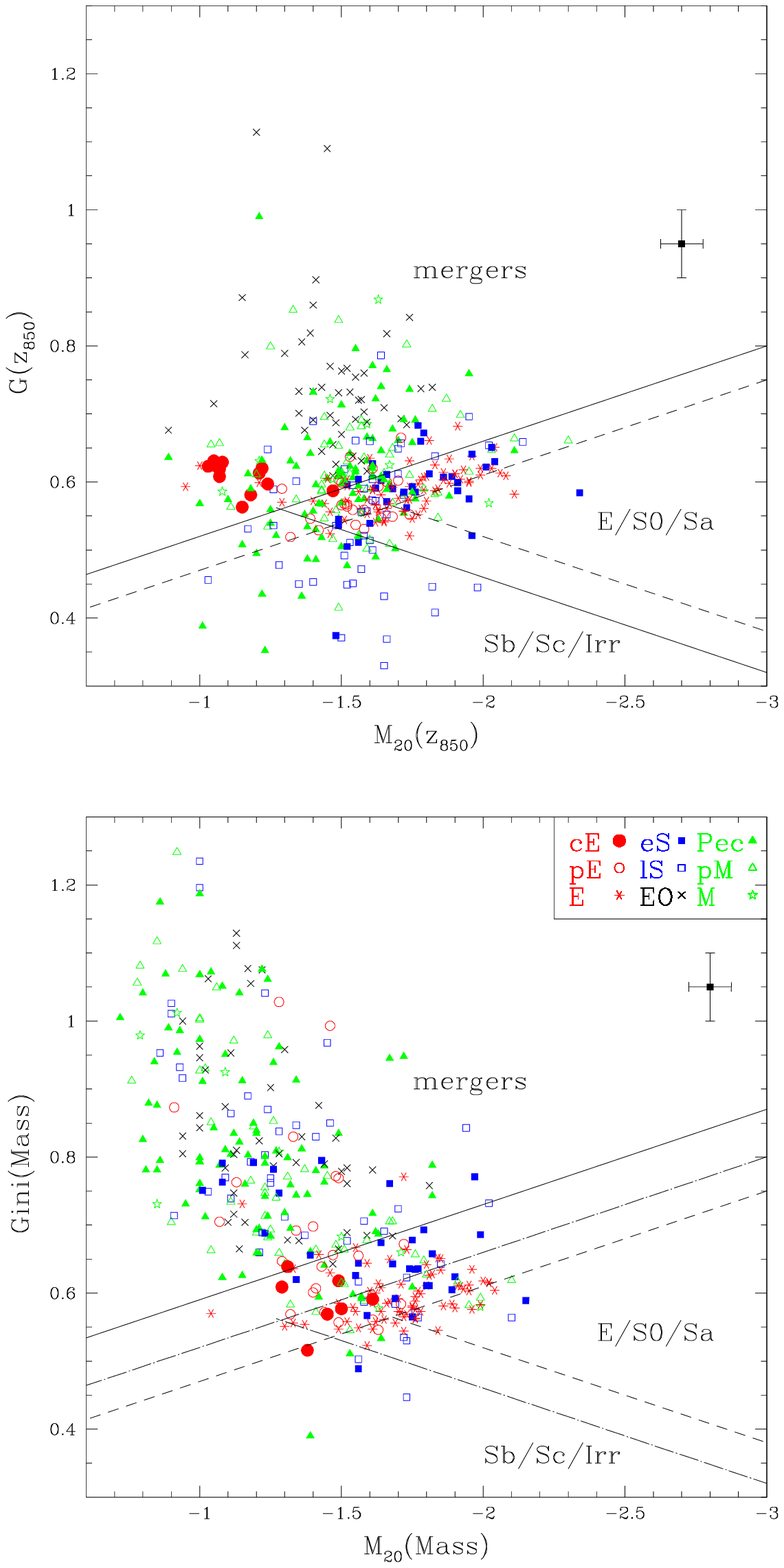}
\caption{Gini-M$_{20}$ for $z_{850}$ (top) and the stellar mass maps 
(bottom). In $z_{850}$ (top) the dashed lines are the classification 
criteria of Lotz \etal  (2004; Equations ~\ref{eq:lotz_mergerline} and
  ~\ref{eq:lotz_normalline}) while the solid lines mark our revised criteria
  in $z_{850}$ (Equations ~\ref{eq:me_mergerline} and
  ~\ref{eq:me_normalline}). In stellar mass (bottom) the Lotz \etal\,  
criteria are
  marked by the dashed lines, our $z_{850}$ criteria are shown by the
  dot-dashed lines and the solid line shows our separation of early and late
  types/mergers, in stellar mass.}
\label{fig:gm20_mass}
\end{figure}

\begin{figure}
\centering
\includegraphics[angle=0, width=90mm, height=90mm]{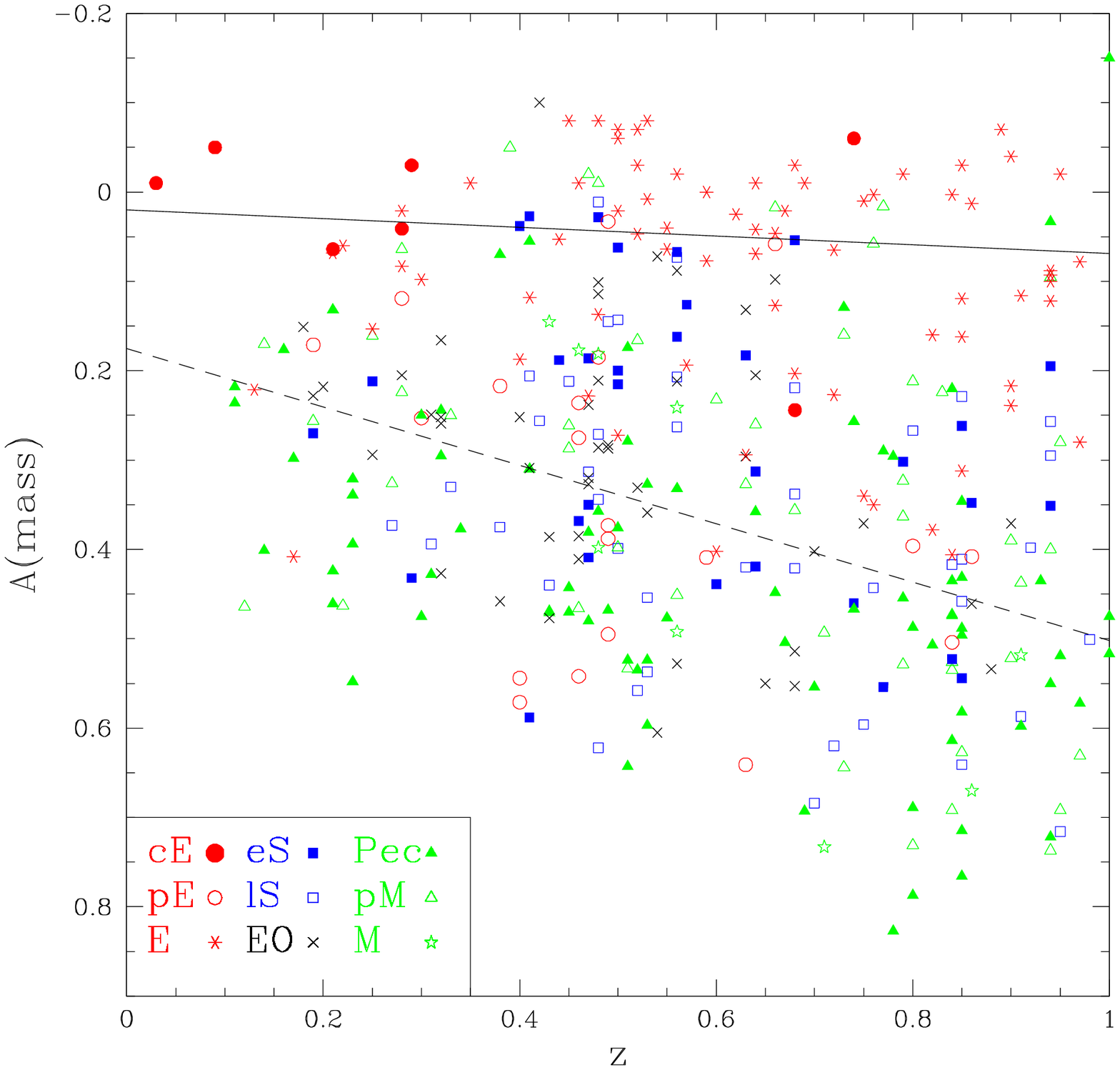}
\caption{Asymmetry in stellar mass maps vs. redshift. The solid line shows the 
fit to the E galaxies (Eq. ~\ref{eqn:E_Am_z}), which is approximately 
constant across redshifts ($z$).  The dashed line illustrates the fit to 
the late-type spiral galaxies (Eq. ~\ref{eqn:lS_Am_z}). The late-type 
spiral galaxies show a larger $A(M_{*})$ with increasingly higher redshifts.}
\label{fig:Am_z}
\end{figure}

Figure ~\ref{fig:AC_mass} shows the $A$-$C$ relation for the stellar mass 
maps. There is a greater range in asymmetry values in stellar mass and the 
criterion suggested by Conselice (2003) for merging systems, of $A > 0.35$, 
even includes a few E and some eS galaxies. There is, however, still a clear 
separation in types between early and peculiar type systems, but the spirals 
(especially lS galaxies) significantly overlap between the two. This makes 
it difficult to distinguish between morphological types in stellar 
mass within $A-C$.  However, given the fact that many of these spirals appear
to have some kind of merger or formation mode, the stellar mass structure
is superior for finding galaxies in active evolution.

\subsubsection{Gini vs. M$_{20}$}

We plot the relation between $G$ and M$_{20}$ for light and stellar mass in 
Figures ~\ref{fig:gm20_bviz} and ~\ref{fig:gm20_mass}. The solid lines in Fig
~\ref{fig:gm20_bviz} are taken from Lotz \etal (2008), who used these
parameters to
classify their sample of normal galaxies and ULIRGs. Equation
~\ref{eq:lotz_mergerline} describes the line above which Lotz \etal classify
their sample as mergers, and Eq. ~\ref{eq:lotz_normalline} describes the line
separating early and late-type galaxies. 

\begin{equation}
G = -0.14 \cdot M_{20} + 0.33
\label{eq:lotz_mergerline}
\end{equation}

\begin{equation}
G = 0.14 \cdot M_{20} + 0.80
\label{eq:lotz_normalline}
\end{equation}

We again see that the cleanest, in terms of separating morphological types, optical
relation is in $z_{850}$, although our sample is not best represented by the
Lotz \etal  classifications in Gini-M$_{20}$ space, with many early-type
galaxies falling into the merging region of the plot. This is partially due to
the fact that $G$ and M$_{20}$ are calculated differently here then in Lotz
\etal, as described in \S3.3 and Lisker \etal (2008).  However, we cannot
rule out that the classifications for this sample may be intrinsically
different, and it is worrying  that all of the edge-on galaxies fall into the
merging region of $G$-M$_{20}$.    We also use a similar
procedure to look at the same rest-frame wavelength as Lotz
et al. does. 

We modify the Lotz et al. (2008) relations for $z_{850}$ (top) and 
stellar mass (bottom) and show this in Fig. ~\ref{fig:gm20_mass}. The 
solid lines in the $z_{850}$
plot are our revisions to the Lotz \etal relations, based on our sample
(Equations ~\ref{eq:me_mergerline} for mergers and ~\ref{eq:me_normalline}
for normal systems),

\begin{equation}
G(z_{850}) = -0.14 \cdot M_{20}(z_{850}) + 0.38
\label{eq:me_mergerline}
\end{equation}

\begin{equation}
G(z_{850}) = 0.14 \cdot M_{20}(z_{850}) + 0.74
\label{eq:me_normalline}
\end{equation}

\noindent while the dashed lines are the original Lotz \etal relations. In
the bottom panel of Fig. ~\ref{fig:gm20_mass}, the solid line shows our
revision of the merger defining line for the stellar mass maps,
(Eq. ~\ref{eq:mass_mergerline}), 

\begin{equation}
G(M_{*}) = -0.14 \cdot M_{20}(M_{*}) + 0.45,
\label{eq:mass_mergerline}
\end{equation}

\noindent while here our relations for $z_{850}$ and
those of Lotz \etal are shown as dot-dashed and dashed, respectively.

As with the relation for $A$($M_{*}$)-$C$($M_{*}$), the 
$G$($M_{*}$)-M$_{20}(M_{*})$ relation shows
the same general trend as its $z_{850}$ counterpart, but with a larger spread
in values. The spirals, again, are not as neatly defined as in $z_{850}$, with
the eS galaxies separated from the early-types and the late-type
spirals and edge-on disk galaxies
occupying the merging region. However, there is a clear separation in
the stellar mass Gini vs. $M_{20}$ which does not exist for the optical
light.  Several of the pE galaxies lie in the merger
region, even as defined by the higher stellar mass line
(Eq. ~\ref{eq:mass_mergerline}). This has also been noted by Conselice \etal
(2007), and is likely due to these objects having multiple nuclei.

The $G$($M_{*}$)-M$_{20}(M_{*})$ relation shows a clear early/late-type 
split, defined by Eq. ~\ref{eq:mass_mergerline}, as the Sb/Sc/Irr region, as 
described by
Lotz \etal is not appropriate to apply to our sample in
stellar mass.  We discuss other correlations: asymmetry vs. clumpiness,
concentration vs. M$_{20}$, size vs. concentration, and asymmetry
vs. M$_{20}$ in the appendix A.  We evaluate in this appendix
other scaling relationship
between these non-parametric relationships, including asymmetry-clumpiness,
concentration and M$_{20}$, size and concentration, and asymmetry and
M$_{20}$.

\subsection{The Evolution of Galaxy Stellar Mass Structure with Redshift}

In this section we investigate the changes in our galaxy stellar mass
maps with redshifts. We take a quantitative approach here and find
that at $z < 1$ there is little change in the stellar mass
structure with redshift, with the exception of the asymmetries of
these galaxies which decreases for some galaxy types at $z < 1$.

We plot $A(M_{*})$ against redshift (Figure
~\ref{fig:Am_z}) where we can see a general trend for
galaxies to become more asymmetric in stellar mass at
higher redshifts.   
We find that not all galaxy types follow this pattern
however, with the most notable exception being the
ellipticals.  As shown in Fig. ~\ref{fig:Am_z}
there is no trend for the ellipticals 
to become more asymmetric in stellar mass at higher
redshifts.

We note that while the Es remain at approximately the same
low asymmetries across the range in redshift, there is a trend for
spiral galaxies to become increasingly asymmetric at higher redshifts. We
measure this trend by fitting both the ellipticals (solid line) and lS (dashed line)
galaxies, plotted in Figure ~\ref{fig:Am_z} and quantified in Equations
~\ref{eqn:E_Am_z} and ~\ref{eqn:lS_Am_z}, below.

\begin{equation}
A_{E}(M_{*}) = 0.049(\pm0.079) \cdot z + 0.021(\pm0.056)
\label{eqn:E_Am_z}
\end{equation}

\noindent is the best fit for the ellipticals and for the late-type spirals,
the best fit is,

\begin{equation}
A_{lS}(M_{*}) = 0.327(\pm0.093) \cdot z + 0.175(\pm0.064).
\label{eqn:lS_Am_z}
\end{equation}

\noindent Equation ~\ref{eqn:E_Am_z} shows that the early-type 
galaxies retain
approximately constant asymmetries across $0 < z < 1$. Equation
~\ref{eqn:lS_Am_z} indicates that there is a relation for late-type spirals,
such that $A(M_{*})$ increases with redshift.


We have already discussed in \S 4.3.3 how the late-type spiral galaxies
have stellar mass asymmetries larger than their asymmetries in
optical light.   Since the late-type spirals are galaxies
which have arms/disks that are larger/brighter than
the bulge component, these galaxies are therefore
dominated in terms of their light by their spiral
structures and arms.  These galaxies show the
most diversity and contract between spatial distributions
within their M/L and stellar mass maps 
(e.g., Fig. ~\ref{fig:massmap_lS}).



We note that not all of the late-type spirals have high asymmetries
in the stellar mass maps, some are as asymmetric or less than in the
$z_{850}$-band (Fig. ~\ref{fig:asym_cf}).  Furthermore, we find that
in general the disk galaxies with higher stellar mass map 
asymmetries have a bluer
colour, which is the case at both high
and low redshifts.   In
many of these cases, the structures are lopsided and/or contain
outer features that remain after normalising by the M/L map.
This interpretation of stellar mass being more localised than the light
is consistent with a higher Gini index in the stellar mass
bands than in the light.

Some of these asymmetric spirals, as well as many of the peculiars
resemble the so-called ``clump
clusters'' and clumpy spirals found by Elmegreen \etal (2005, 2007), 
especially in stellar mass maps.  Their low concentration in mass
is also similar to the Luminous Diffuse Objects (LDOs) found
at similar redshifts (Conselice et al. 2004).  
Indeed, there are some late-type spirals galaxies that we would
have classified differently, perhaps as Pec or pM systems, had we performed
the classification in stellar mass, rather than in $z_{850}$. 
The clumps in galaxies
found by Elmegreen \etal are also massive, with typical values in the region
$\sim10^{8} - 10^{9}M_{\odot}$ (Elmegreen \& Elmegreen 2005), and reside in
young disks at high redshift. It is likely that some of our highly asymmetric
late-type spiral galaxies are similar, or in the same class as these clumpy 
spirals.  These features may also be the result of minor merger events.  

\section{Conclusions}\label{sec:conclusions}

We conduct a pixel by pixel study of the structures of 560 
galaxies found in the 
GOODS-N field at $z < 1$ within stellar mass maps and Hubble
Space Telescope ACS $BViz$ wavebands.
We measure stellar masses for each pixel of each galaxy image from our
$BViz$ images by fitting to stellar population models. We use these 
values to construct stellar mass maps for our sample and compare morphologies 
in $BViz$ and stellar mass. Our major findings and results include:

I. We construct stellar mass maps and mass-to-light ratio maps for each 
galaxy and we present examples of each morphological type, in the 
$z_{850}$ band, stellar mass maps, and $(M/L)_B$ in 
Figures ~\ref{fig:massmap_cE} to ~\ref{fig:massmap_M}. We
find that some compact elliptical (cE) galaxies have blue cores, and more
complicated internal structures than in either $z_{850}$ or stellar mass would 
reveal on their own.  Many of the peculiar
ellipticals (pE) also have blue cores and we assert that these objects may
have merged in their recent histories.
The early and late-type spirals display more varied patterns. We find that for
some of these galaxies, structures seen in light (e.g. spiral arms) are
smoothed out in stellar mass. However, this effect does not hold true
for all galaxies, nor for all features seen in light.
Stellar mass and ($M/L$) maps of the `peculiar' galaxies are complicated and 
vary across the sample. Although it is more difficult to make out features in 
stellar mass for galaxies that have very blue colours, 
structures not seen in light are revealed.

II. We compare the half-light
radius, $R_{\rm e}$, in $z_{850}$ with half-mass radii for our sample
as a function of morphological type. We find no systematic tendency 
for any particular morphological type to have larger $R_{\rm e}$ in stellar mass 
than in the  $z_{850}$ band, and
thus conclude that there is no bias introduced by measuring galaxy sizes in
$z_{850}$.

III. We find a clear tendency for many galaxies to be more asymmetric
in stellar mass than in $z_{850}$. We also find that morphology correlates
with asymmetries, with early-types having low $A(M_{*})$, and late-types 
higher values of $A(M_{*})$. We find a relation between colour and 
$A(M_{*})$ such that bluer galaxies have higher stellar mass asymmetry
differences between optical light and stellar mass maps. 
The late-type 
spiral galaxies in the sample have higher $A(M_{*})$ than 
would be expected from their asymmetries in $z_{850}$. We discuss possible 
causes of this effect, including regions of enhanced star formation also 
possessing higher stellar masses, and the evolution of spirals. We note that these highly asymmetric
spirals resemble the clumpy disks of Elmegreen \etal (2007), and Conselice et al. (2004) 
and are experiencing either minor merging activity, or bulge formation
through accretion of disk material.

IV. We find that the Gini index in stellar mass is higher than in the
z-band ($G(M_{*}) > G(z_{850})$) for all galaxies except the early-types,
indicating that most of the stellar mass in later morphological types is 
contained within fewer pixels. Late-type and peculiar morphologies show a 
trend for M$_{20}(M_{*})$ $>$ M$_{20}(z_{850})$, suggesting that the 
brightest $20$ percent of the brightest pixels is not necessarily where 
the greatest $20$ percent of the stellar mass is located.

V. We investigate the relations between several combinations of the
$CAS$ parameters in $BViz$ and stellar mass, and compare our 
results to previous morphological studies of this type (e.g. Conselice, 
Rajgor \& Myers 2008; Lotz \etal  2004; 2008). We find that $z_{850}$ is 
the most appropriate photometric band to utilise for a $z < 1$ sample of 
galaxies with all morphologies, although stellar mass maps are better at
distinguishing active galaxies from passive ones and is more a physical
measure of structure.


We furthermore compare our sample classifications in $G$-M$_{20}$ to those of Lotz \etal
(2008) and find that the Lotz \etal criteria do not best describe our
sample. We revise the Lotz \etal (2008) criteria to best fit our sample
in the $z_{850}$-band and within the stellar mass maps, and find that 
early-types, late-types and mergers can be separated. However, the edge-on
disk galaxies remain problematic and cannot be distinguished from mergers in
$G(z_{850})$-M$_{20}(z_{850})$. We find that $G(M_{*})$-M$_{20}(M_{*})$ 
can be used to broadly separate early from late-type galaxies, but the 
criteria cannot distinguish between late-type/edge-on disks 
from peculiar/merger systems.

VI. We find a relationship between $R_{\rm e}$ and $C$ (see appendix A) 
for early-type galaxies in both
$z_{850}$ and stellar mass. In each case we find that galaxies with higher
concentrations have larger radii, and this relation is steeper in $z_{850}$
than stellar mass. We also investigate asymmetry versus M$_{20}$ and find that
these parameters display a similar relation to $A-C$ in $i_{775}$ and
$z_{850}$. In stellar mass, $A(M_{*})$-M$_{20}(M_{*})$ shows a tighter relation 
than $A(M_{*})-C(M_{*})$, with a clear separation between early and late-type
systems. Thus, we conclude that, for all parameters, late-type
spiral galaxies overlap with with Pec/pM/M systems which may be
ultimately taken from the same subset of galaxies. Structural studies in 
stellar mass do however track both minor and major merging events, and
can be used to find galaxies in active galaxy evolution modes.

We thank the GOODS team for making their data public, and STFC for a studentship towards supporting this work
and the Leverhulme Trust for support.

\appendix

\section{Appendices}

\section{Structure Scaling Relations}

In this appendix we describe several additional scaling relationships between structural 
and morphological parameters, not discussed in the body of the paper, 
for our galaxy sample.  Many of the relations
we see here confirm the conclusions we reached earlier. Furthermore,
we list these for a complete overview of how structure correlates in
the Hubble Space Telescope ACS $BViz$ bands as well as in stellar mass maps. 

\begin{figure*}
\centering
\includegraphics[angle=0, width=148mm, height=148mm]{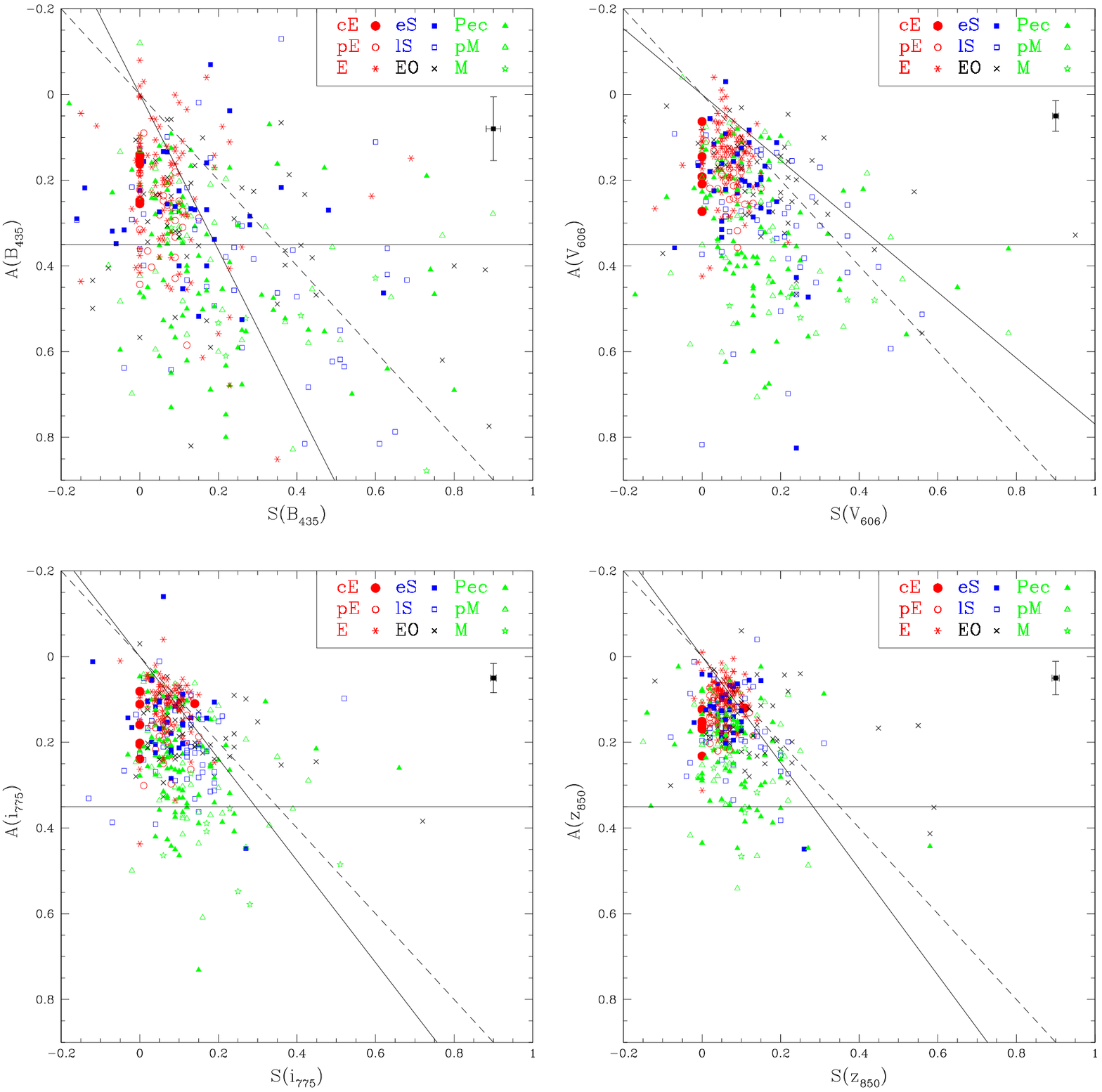}
\caption{Asymmetry vs. clumpiness for the $B_{435}$, $V_{606}$, $i_{775}$ and
  $z_{850}$ bands. The solid horizontal line marks $A = 0.35$. The solid
  diagonal line shows the fit for each band to the elliptical and
spiral (i.e., `normal') galaxies in
  the sample and the dashed line marks the position of $A = S$. The black
  square point in the top right hand section of the plot illustrates the
  average error for the whole sample.}
\label{fig:AS_bviz}
\end{figure*}

\subsection{Asymmetry vs. Clumpiness}\label{sec:AS}

\begin{figure}
\centering
\includegraphics[angle=0, width=75mm, height=75mm]{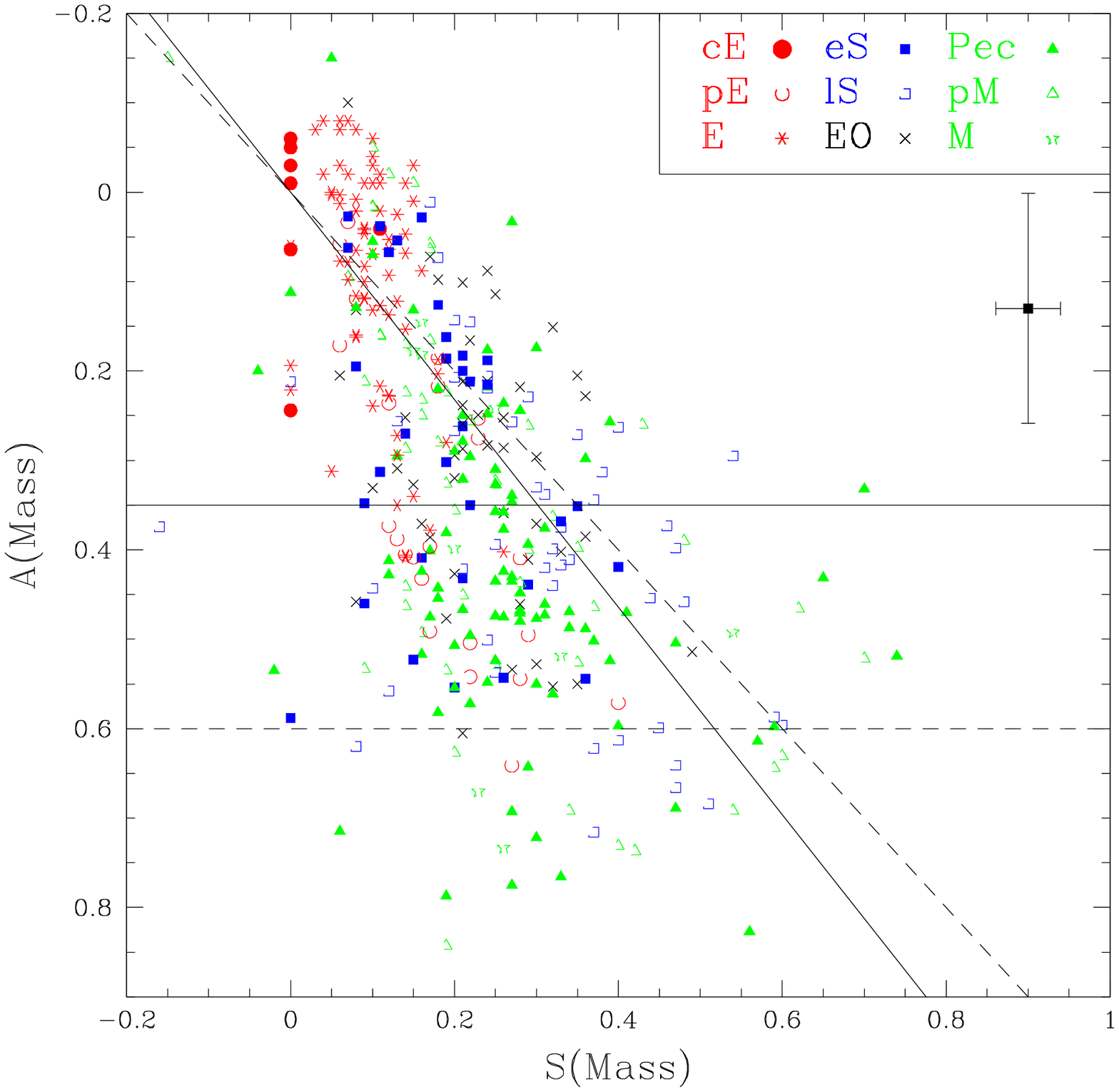}
\caption{Asymmetry vs. clumpiness for the stellar mass maps. The convention is the same
as in Fig. ~\ref{fig:AS_bviz}. The horizontal dashed line marks the position
of $A = 0.6$, which we derive as a criterion for merger
candidates in stellar mass.}
\label{fig:AS_mass}
\end{figure}

We postulate that the many small scale asymmetries, measured by the
clumpiness parameter, must approximate to the asymmetry of the galaxy as a
whole. Figures ~\ref{fig:AS_bviz} and ~\ref{fig:AS_mass} show the relation
between $A$ and $S$ for $B_{435}$, $V_{606}$, $i_{775}$, $z_{850}$ and  stellar
mass,
respectively. The horizontal solid line in each plot marks the $A =  0.35$
position and the dashed line in each plot illustrates where $A = S$.  The
solid diagonal lines in each plot are the fits to the E, eS and lS  galaxies
in each band, which were forced to coincide at $A=S=0$. These fits are shown
below in Equations  ~\ref{eqn:AS_B}-~\ref{eqn:AS_mass}, for each band
respectively.

\begin{equation}
A(B_{435}) = 0.55(\pm0.07) \cdot S(B_{435})
\label{eqn:AS_B}
\end{equation}

\begin{equation}
A(V_{606}) = 1.30(\pm0.08) \cdot S(V_{606})
\label{eqn:AS_V}
\end{equation}

\begin{equation}
A(i_{775}) = 1.20(\pm0.08) \cdot S(i_{775})
\label{eqn:AS_i}
\end{equation}

\begin{equation}
A(z_{850}) = 1.24(\pm0.08)\cdot S(z_{850})
\label{eqn:AS_z}
\end{equation}

\begin{equation}
A(M_{*}) = 1.16(\pm0.06) \cdot S(M_{*})
\label{eqn:AS_mass}
\end{equation}

\noindent It is evident that the relations are mostly steeper than $A = S$, although the 
fit in stellar mass comes closest to this equality. 
The $A$($M_{*}$)-$S$($M_{*}$) relation thus is useful for identifying merger
candidates. The criterion of $A$($M_{*}$) $> 0.35$ is too low for the stellar mass maps,
instead we suggest that galaxies which lie in the region above $A$($M_{*}$) $>  0.6$
(horizontal dashed line, Fig. ~\ref{fig:AS_mass}) and underneath the
$A$($M_{*}$)-$S$($M_{*}$) fit, be classified as mergers, and this  could be adapted for
use to calculate merger fractions across a range of  redshifts.

\begin{figure*}
\includegraphics[angle=0, width=148mm, height=148mm]{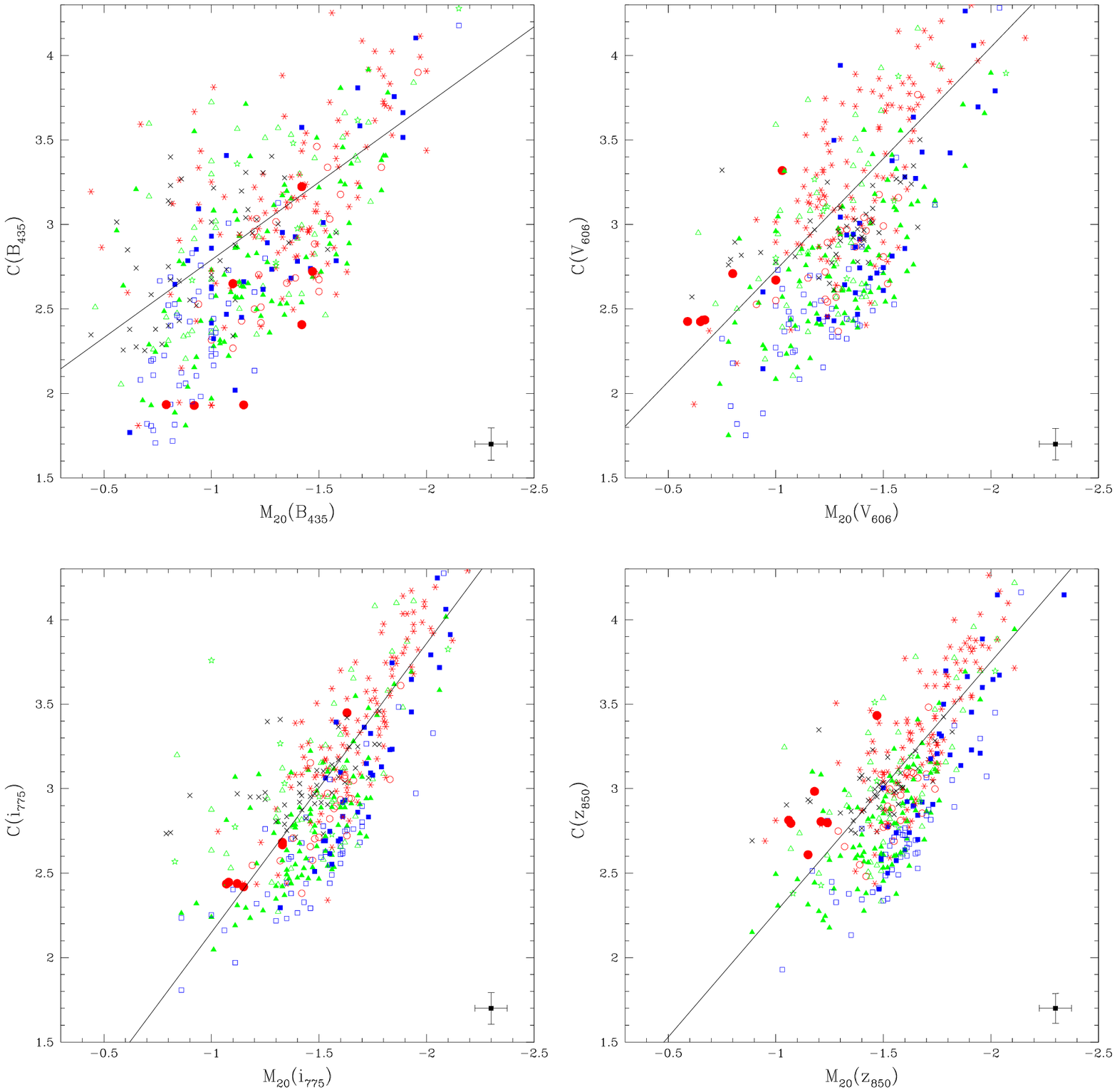}
\caption{The relation between C-M$_{20}$ for the $B_{435}$, $V_{606}$, $i_{775}$ and $z_{850}$
bands. The solid lines show our fits to the E and eS galaxies in each band
(Equations ~\ref{eqn:CM20_B} to ~\ref{eqn:CM20_z}). The points are the same as in Figure~A1.}
\label{fig:cm20_bviz}
\end{figure*}

The trend noted  for the parameters to be less reliable 
in the bluer bands also holds here (Fig ~\ref{fig:AS_bviz}). We note that 
the criteria $A(z_{850}) > 0.35$ and $A(z_{850}) > 1.24\cdot S(z_{850})$ are 
appropriate for identifying merger candidates, and calculating merger 
fractions. The fits to $A$-$S$ for $i_{775}$, $z_{850}$ and stellar mass are very 
similar, despite the much larger scatter in $A$($M_{*}$).

\begin{figure}
\includegraphics[angle=0, width=75mm, height=75mm]{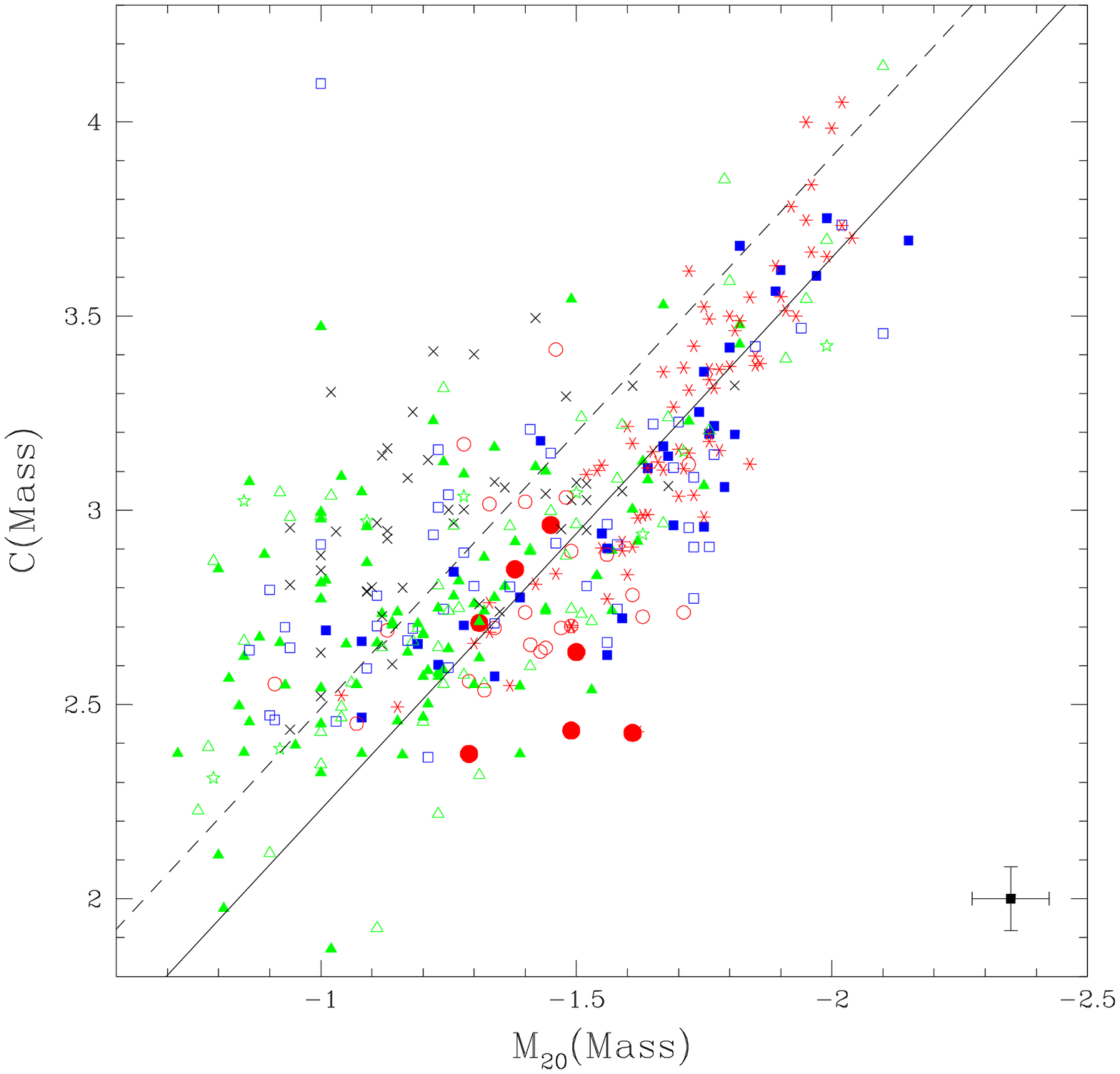}
\caption{The relation between C-M$_{20}$ for the stellar mass maps. The solid line 
shows our fit to the E
  and eS galaxies in stellar mass (Eq. ~\ref{eqn:CM20_mass}). The solid line is this 
relation plus $2\sigma$. The points are the same as in Figure~A1.}
\label{fig:cm20_mass}
\end{figure}

\subsection{M$_{20}$ vs. Concentration}

In this section we investigate the relationship between M$_{20}$ and $C$ in 
$B_{435}$, $V_{606}$, $i_{775}$, $z_{850}$ and stellar mass (Figures 
~\ref{fig:cm20_bviz} and ~\ref{fig:cm20_mass}). The solid lines in each 
plot show the fits to the E and eS galaxies in the sample. The forms of 
these fits are shown below in Equations ~\ref{eqn:CM20_B} to 
~\ref{eqn:CM20_mass}.

\begin{equation}
C(B_{435}) = -0.92(\pm0.12) \cdot M_{20}(B_{435}) + 1.87(\pm0.16)
\label{eqn:CM20_B}
\end{equation}

\begin{equation}
C(V_{606}) = -1.32(\pm0.14) \cdot M_{20}(V_{606}) + 1.41(\pm0.20)
\label{eqn:CM20_V}
\end{equation}

\begin{equation}
C(i_{775}) = -1.71(\pm0.13) \cdot M_{20}(i_{775}) + 0.44(\pm0.22)
\label{eqn:CM20_i}
\end{equation}

\begin{equation}
C(z_{850}) = -1.48(\pm0.13) \cdot M_{20}(z_{850}) + 0.79(\pm0.22)
\label{eqn:CM20_z}
\end{equation}

\begin{equation}
C(M_{*}) = -1.42(\pm0.08) \cdot M_{20}(M_{*}) + 0.81(\pm0.13)
\label{eqn:CM20_mass}
\end{equation}

M$_{20}$ is anti-correlated with $C$, such that highly concentrated 
galaxies have lower M$_{20}$ values. The general trend for 
concentration to increase as M$_{20}$ decreases holds for all bands, although
the scatter increases at higher values of M$_{20}$ and lower $C$. The 
relation has a larger dispersion in the $B_{435}$ and $V_{606}$ bands, 
but this is more ordered than the random scatter seen in previous 
relations (Fig. ~\ref{fig:cm20_bviz}, top panels). The relation is 
distinct in both $i_{775}$ and $z_{850}$, with the early-type galaxies 
occupying the high 
$C$, low M$_{20}$ region, the eS's having higher M$_{20}$ values for their 
concentration and late-types generally having low $C$ and high M$_{20}$.

This trend is also seen in the stellar mass maps (Fig. ~\ref{fig:cm20_mass}), but
with greater separation between the morphological types than in $i_{775}$ and
$z_{850}$.  The early-types also have high concentration and low M$_{20}$,
although  the  spirals are more randomly distributed, and do not occupy any
one region of  the plot. The late/Pec/pM/M galaxies have higher M$_{20}$
values for their  concentrations. The dashed line in Fig. ~\ref{fig:cm20_mass}
shows the fit  plus $2\sigma$ and we postulate that galaxies falling above
this criterion  can be picked out as merger candidates.

\subsection{Size vs. Concentration}

\begin{figure}
\centering
\includegraphics[angle=0, width=90mm, height=180mm]{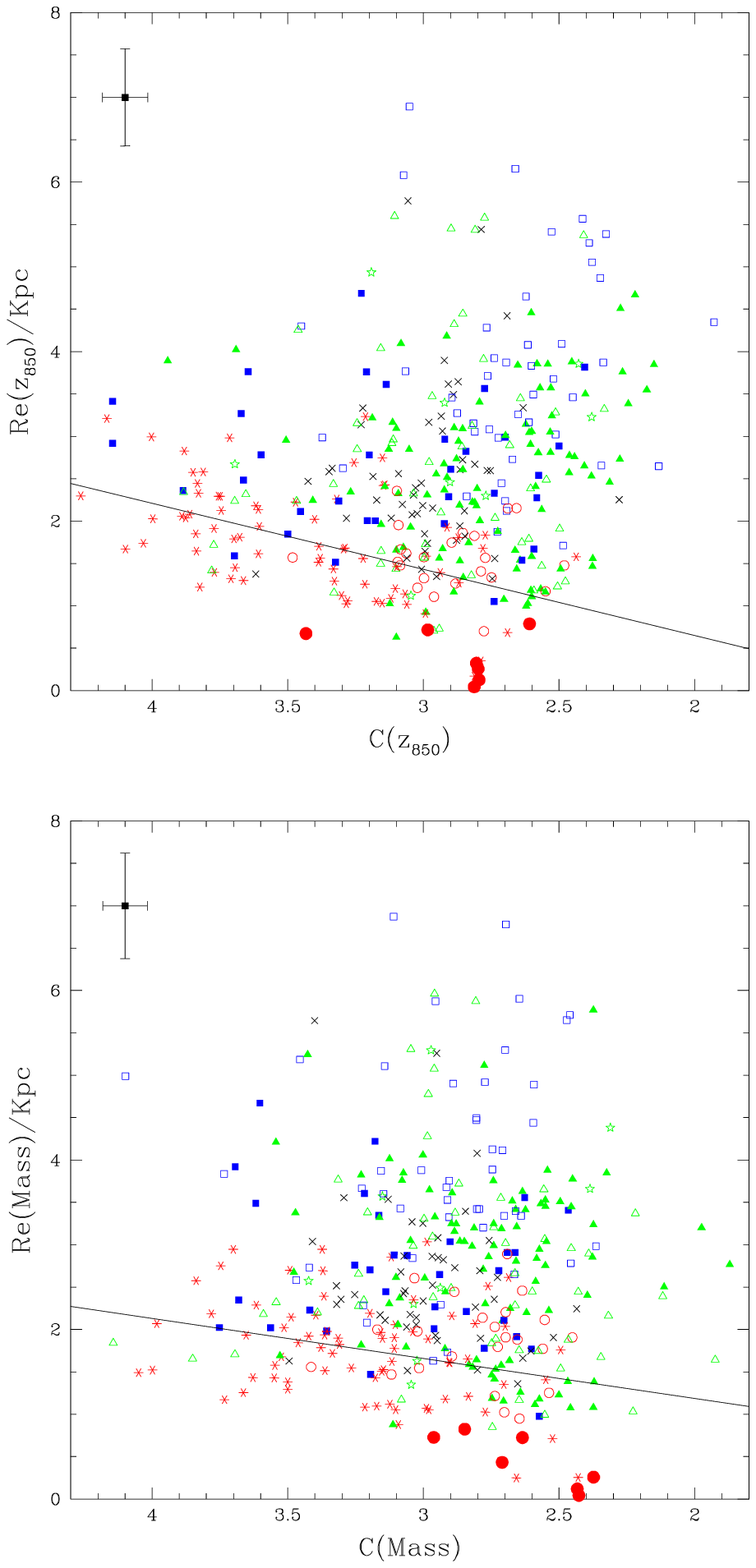}
\caption{The relation between $R_{\rm e}$ and concentration for 
$z_{850}$-band images (top) and stellar mass maps (bottom). The solid lines 
show the
fits to the E galaxies in each band (Equations ~\ref{eqn:ReC_z} and ~\ref{eqn:ReC_mass}). The points are the same as in Figure~A1.}
\label{fig:rec_zm}
\end{figure}

In this section we investigate the relation between galaxy size, as 
measured by the half-light radius, $R_{\rm e}$, and concentration, 
in $z_{850}$ and stellar mass. The top panel in Fig. ~\ref{fig:rec_zm} 
shows the relation in 
$z_{850}$ and the bottom panel in stellar mass. We find that the early-types 
follow a linear relationship, and we have fit this relation for the E 
galaxies in $z_{850}$ (Eq. ~\ref{eqn:ReC_z}) and stellar mass (Eq. 
~\ref{eqn:ReC_mass}). 

\begin{equation}
Re(z_{850}) = 0.78(\pm0.15) \cdot C(z_{850}) - 0.91(\pm0.52)
\label{eqn:ReC_z}
\end{equation}

\begin{equation}
Re(M_{*}) = 0.48(\pm0.21) \cdot C(M_{*}) + 0.23(\pm0.68)
\label{eqn:ReC_mass}
\end{equation}

\noindent The relation in $z_{850}$ is steeper than in stellar mass, and both 
trends 
show that as $R_{\rm e}$ increases, $C$ also increases. For normal galaxies, $C$ 
is a tracer of stellar mass, such that lower $C$ corresponds to less massive 
galaxies (Conselice 2003). In both $z_{850}$ and stellar mass we see that 
more concentrated galaxies have larger radii, and this effect is more 
pronounced in $z_{850}$ than in stellar mass. No such relation is found 
for the late-types and Pec/pM/M systems. 

Conselice \& Arnold (2009) examine galaxies at $z = 4-6$ in the 
Hubble Ultra Deep Field (HUDF) finding that early-types 
are smaller for their stellar mass at higher redshifts. They conclude that those 
galaxies that follow the $C-R_{\rm e}$ relation are in a relaxed state.

\begin{figure*}
\centering
\includegraphics[angle=0, width=138mm, height=138mm]{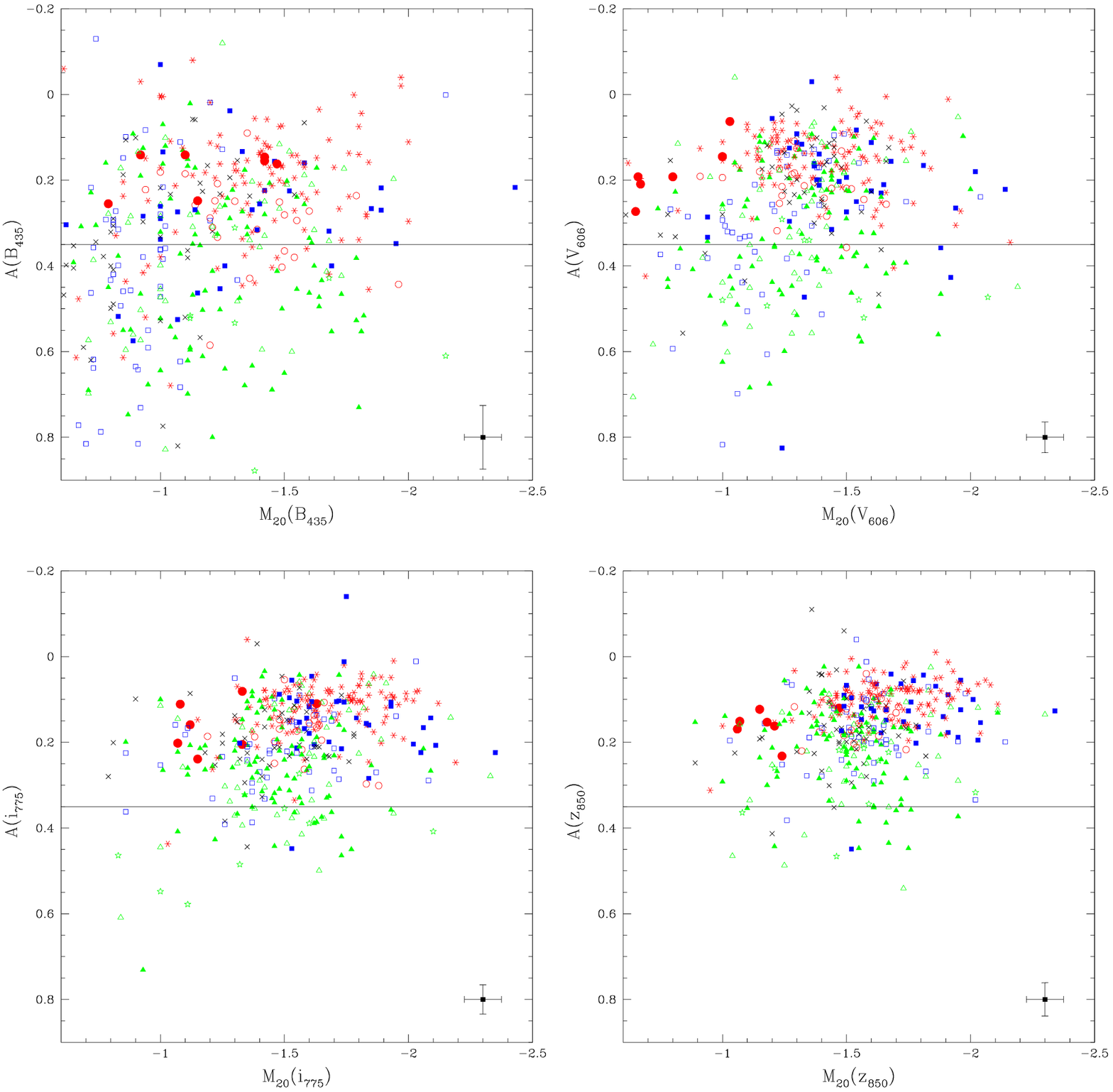}
\caption{Asymmetry-M$_{20}$ for the $B_{435}$, $V_{606}$, $i_{775}$ and
  $z_{850}$ bands. The horizontal solid line marks $A = 0.35$ in each band. The points 
are the same as in Fig. A1.}
\label{fig:AM20_bviz}
\end{figure*}

\subsection{Asymmetry vs. M$_{20}$}

\begin{figure}
\centering
\includegraphics[angle=0, width=75mm, height=75mm]{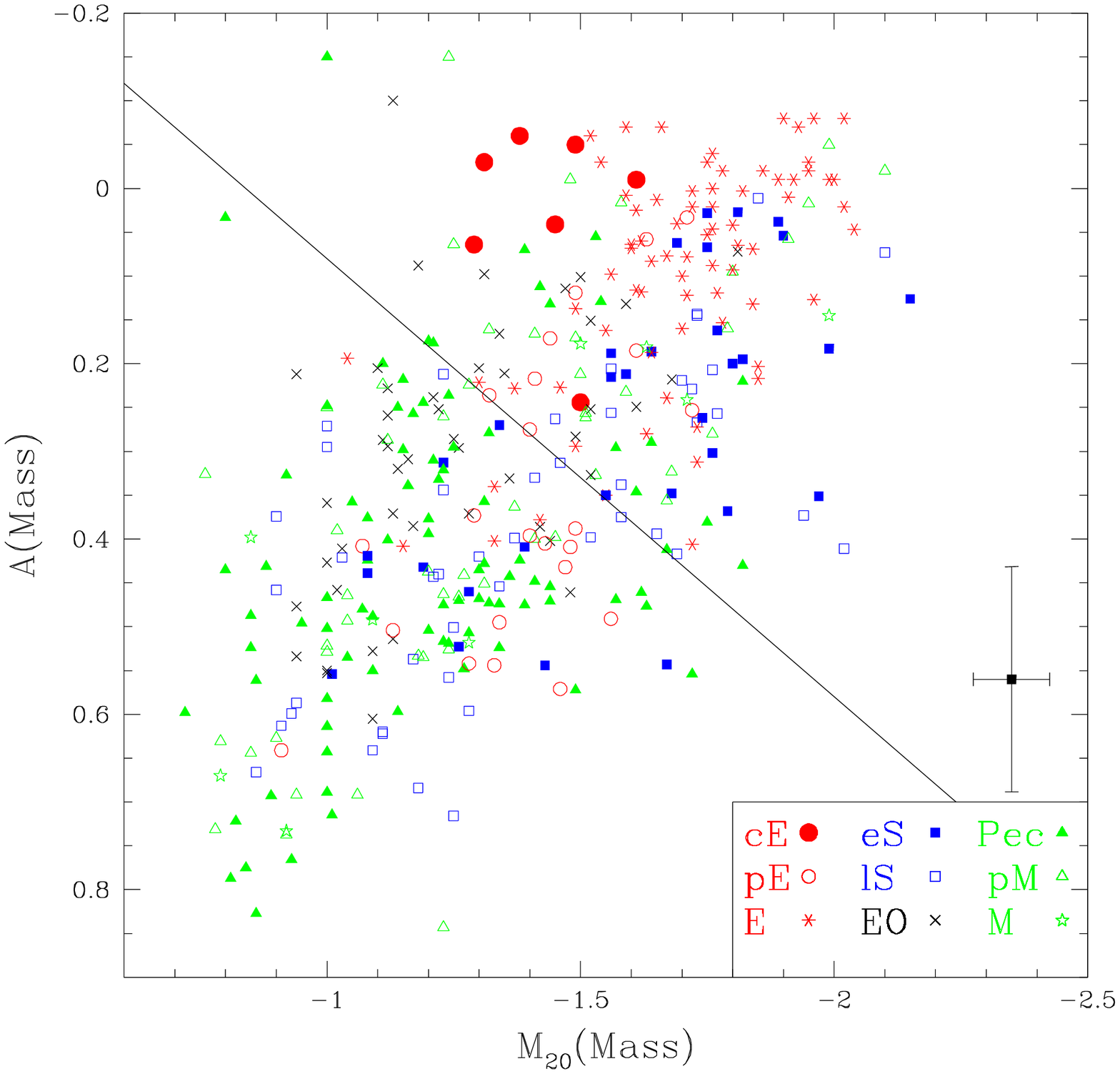}
\caption{Asymmetry-M$_{20}$ for the stellar mass maps. The solid diagonal line marks
  our division between early and late-types/mergers.}
\label{fig:AM20_mass}
\end{figure}

Here we investigate the relation between asymmetry and M$_{20}$ in  the $B_{435}$,
$V_{606}$, $i_{775}$, $z_{850}$ bands (Fig. ~\ref{fig:AM20_bviz}) and  stellar mass
(Fig. ~\ref{fig:AM20_mass}). As M$_{20}$ is analogous to $C$, we  would expect
to see a similar relation between $A-M_{20}$ and $A-C$. In
Fig. ~\ref{fig:AM20_bviz} we see that in the  $BViz$ bands we indeed find a 
similar
relation in $A-M_{20}$ to $A-C$, and could be used alongside $A-C$
for classification purposes.

In stellar mass, however, the relation between $A-M_{20}$ appears tighter than
$A(z_{850})-C(z_{850})$, such that galaxies with higher asymmetries have lower
values  of $M_{20}$. There is a clear separation between early and late-type
galaxies in stellar mass, and we have included an approximate divide in this plot
(solid line in Fig. ~\ref{fig:AM20_mass}), the form of  which is shown below
in Equation ~\ref{eq:AM20_mass_div}.

\begin{equation}
A(M_{*}) = -0.5 \cdot M_{20}(M_{*}) - 0.42
\label{eq:AM20_mass_div}
\end{equation}

\noindent Although this divide is very successful in separating early-types from 
Pec/pM/M systems, the spirals are spread over these two regions, 
again implying that measuring structural parameters in stellar mass does 
not distinguish between morphologies as well, although it traces physical
processes better.

However, it is possible that viewing the stellar mass distribution of galaxies reveals
different physical processes to those we expect to see in light. We have
already discussed the possibility that some highly asymmetric late-type
spirals in our sample could be in the process of forming bulges. 
It is  also possible that studies in stellar mass could
provide clues as to the formation mechanism involved. The asymmetric 
late-type spirals, for example object 19535 (Figs. ~\ref{fig:massmap_lS},
~\ref{fig:spirals_z}), could be undergoing outside-in
formation, perhaps through the accretion of smaller satellites. We could,
therefore, be picking out systems experiencing minor merging events.

\section{Characterisation of Different Galaxy Types}

At the start of this paper we describe how we classified each of
our galaxies into one of nine type (\S 3.2).  In this section,
we described, based on our stellar mass map analysis, 
the properties of each of these galaxy types in terms
of their optical and stellar mass structures, and what these
imply about the evolution of different classes of galaxies.

While galaxy classification in terms of overall visual
morphology is limited (Conselice 2006b), it is still useful
for connecting galaxies at different redshifts, and for
understanding how major nearby galaxy types such as disks
and ellipticals are formed.  Below we give details 
on the properties of each of the nine galaxy type that
we have examined within GOODS.
 
\subsection{Compact Es}

Compact ellipticals are typically found in the late-type area of the
C-A optical light diagram, but are within the early-type region of the 
C-A stellar mass diagram.  The clumpiness values for these
systems are also zero in all wavelengths and stellar mass due
to the fact that these systems are so small this parameter cannot
be reliably measured.  The concentrations of these systems
are always quite low, with $C=2.5-3$, at the resolution limit
where we can measure these values, suggesting
that they are not similar to giant ellipticals (e.g.,
Conselice 2003).

Some interesting features appear when we compare the quantitative
properties of cEs as measured in optical light to those in stellar
mass maps.  One of these is that for all cEs, the size measured
in the stellar mass maps are always larger than for those measured
in the reddest band, $z_{850}$.   The asymmetry measured in stellar
mass maps is also quite low, and lower than in the \zband,
perhaps due to the fact that the asymmetric light is brought to a lower
level within the stellar mass images compared to the optical light.
The M$_{20}$ parameter is also quite different between the stellar
mass maps and the \zband, such that M$_{20}$ is higher (less negative) in 
the \zband.  This implies that the light distribution is
more concentrated in the stellar mass maps, likely due to the 
blue coloured core.

\subsection{Ellipticals}

First, we find that just based on colours, there is a wide
diversity at a given redshift (see Figs. 12 and 13) - an indication that 
at higher redshifts there is a range of elliptical galaxy
star formation histories,
something that has been recognised for some time
(e.g., Stanford et al. 2004; Conselice et al. 2007).
Simply based on the C-A diagram with optical light
(Fig. 20) it appears that most of the early-types
are found in the region of C-A space classifiable
as mid-types.  This is commonly seen for high redshift
ellipticals which appear more asymmetric than their
$z = 0$ counterparts (e.g., Conselice et al. 2007).
Ellipticals do have generally lower Gini and M$_{20}$
indices, however there is still some spread in their
values.

The stellar mass maps reveal that ellipticals appear to
become more `diffuse' within stellar mass maps.  However
the sizes are either very similar in stellar mass maps
or even slightly smaller, than in the corresponding 
$z_{850}$-band image.  The Gini and the M$_{20}$ parameters
do not change significantly between the stellar mass maps and the
$z_{850}$-band image.  However, we find that some
of the ellipticals are slightly more asymmetric in 
stellar mass maps than in the $z-$band,
although many of these systems tend to be bluer
galaxies.  The ellipticals
also tend to become less concentrated in stellar mass
compared with the $z_{850}$ band (Fig. ~\ref{fig:conc_cf}).

Finally, the elliptical galaxies are one of the few
types which do not change significantly in terms of
their Gini and M$_{20}$ values between stellar mass
and the $z_{850}$, and this results in these systems
having a well defined location in the stellar
mass Gini/M$_{20}$ diagram (Fig. ~\ref{fig:gm20_mass}).

\subsection{Peculiar Ellipticals}

The peculiar ellipticals are a galaxy type which has
been recognised to exist up to at least $z = 1.4$
before (e.g., Conselice et al. 2007).  These peculiar
ellipticals appear elliptical in overall morphology,
but have internal structure suggesting that they
have recently undergone some type of assembly.
These galaxies are not uncommon, and are more
often seen than the compact elliptical type.

These galaxies are more asymmetric and less concentrated
than the Es - with all having concentrations, $C < 3$.
They are quite distinct from the ellipticals in the
CA plane and are bluer by $\sim 0.5$ magnitudes from the
giant ellipticals.  These blue features can be readily
seen within the M/L maps for these systems 
(Fig. ~\ref{fig:massmap_pE}).  These galaxies often appear more
elliptical in the stellar mass maps once their
M/L variations have been taken into account.

Another difference between the pEs and the Es is that these
systems have half-light (or half-mass) radii which are nearly always
bigger in the stellar mass maps than in the $z_{850}$ band.
This is a sign for star formation activity in the cores
of these galaxies, since the bluer material in the core
is less prominent in the stellar mass maps, giving them
a larger effective radius.  This can also be seen in the
fact that the asymmetry in the stellar mass maps is
generally higher than in the $z_{850}$ band for
these systems, which is another indication that they are
undergoing some form of star formation.

The Gini values for these systems are higher in the
stellar mass band than in $z_{850}$, which demonstrates
that the stellar mass distribution is more concentrated
in bright (but non-central) pixels than the light, as the light is less 
equally distributed than the
stellar mass.  The location of the pEs is similar to that
of the Es in the Gini/M$_{20}$ plane.

\subsection{Spirals}

The spiral galaxies are amongst the most interesting for
this study given that they often contain two major stellar
population types segregated spatially. Traditionally 
this is seen as an older stellar population making up
the bulge or centre, and the spiral arms consisting of
younger stellar populations.  

We find that the early-type spirals, those whose
apparent bulge is brighter/larger than its apparent
disk, are quite blue in colour.  This difference
is very obvious when examining the M/L
ratio maps for these galaxies (Fig. ~\ref{fig:massmap_eS}),
although the differences are not as obvious as with
the late-type spirals.  Nevertheless, these systems
often appear to have a blue outer M/L ratio and a
redder inner one, although there are examples
where this is not the case.  

The sizes of these galaxies does not change significantly
between the stellar mass maps and the $z_{850}$ band
images.  There are however some differences between the
stellar mass maps and the $z_{850}$ band images within
the various parameters.  The most significant of these
is that the asymmetry parameter is higher in the stellar
mass maps than in the $z_{850}$ band imaging as 
discussed in detail in \S 4.3.3.  The Gini
parameter is also higher within the stellar mass
maps than within the $z_{850}$ by roughly
10 percent. This implies that the stellar mass is distributed in fewer
pixels than in the light for spirals.  The distribution of
M$_{20}$ values is similar between the light and stellar mass,
suggesting that the spatial distribution is similar.

One notable feature of the spiral galaxies in general
is that they have a much larger range of asymmetry
values within the stellar mass images.  In fact,
their values are often as high as the peculiar
galaxies, suggesting
that within stellar mass maps these parameters cannot
easily distinguish between mergers and spirals with
mixed stellar populations.

\subsection{Peculiars/Mergers}

The Peculiars, Pre-Mergers and Mergers all have similar patterns
in the CAS space in both the stellar mass maps and within the
$z_{850}$ diagrams.  These are the most asymmetric and bluest
galaxies in our sample, and are likely in some phase of a 
merger, as has been described earlier in e.g.,
Conselice et al. (2003, 2008).

These peculiars (a term we use here for all three of these
types) do not change in size much between the stellar mass and the
$z_{850}$ image.  However, the asymmetry in the stellar mass
band is higher than for the $z_{850}$ band. This is likely
due to the stellar mass within these galaxies being more
concentrated in fewer locations.  The M$_{20}$ also shows
that the light is less concentrated in than the stellar mass, showing
that the stellar mass is more distributed spatially than
in light, suggestive of a merging systems. In fact, these
peculiars have the highest asymmetries, M$_{20}$, and Gini 
indices.

\end{document}